\documentclass[twocolumn,twocolappendix]{aastex702}

\usepackage{amsmath}
\usepackage{comment}
\usepackage{enumitem}
\usepackage{booktabs}
\usepackage{threeparttable}
\usepackage{multirow}

\newcommand{\cxo}{\emph{Chandra}}
\newcommand{\nus}{\emph{NuSTAR}}
\newcommand{\nh}{$N_{\rm H}$}
\newcommand{\cms}{\,cm$^{-2}$}
\newcommand{\ergs}{\,erg\,s$^{-1}$}
\newcommand{\ergcm}{\,erg\,cm$^{-2}$\,s$^{-1}$}
\newcommand{\kms}{\,km\,s$^{-1}$}
\newcommand{\psr}{PSR J1747$-$2958}

\begin{document}
\title{Evidence of a Low-Energy Cutoff in the Injected Spectrum of a Pulsar Wind Nebula}

\author[orcid=0000-0002-1904-4957]{Zhihong Shi}
\affiliation{Department of Physics, The University of Hong Kong, Pokfulam Road, Hong Kong}
\affiliation{Hong Kong Institute for Astronomy and Astrophysics, The University of Hong Kong, Pokfulam Road, Hong Kong}
\email{neoshekzh@connect.hku.hk}

\author[orcid=0000-0002-5847-2612]{C.-Y. Ng}
\affiliation{Department of Physics, The University of Hong Kong, Pokfulam Road, Hong Kong}
\affiliation{Hong Kong Institute for Astronomy and Astrophysics, The University of Hong Kong, Pokfulam Road, Hong Kong}
\email{ncy@astro.physics.hku.hk}

\author[orcid=0000-0002-8848-1392]{Niccol\`o Bucciantini}
\affiliation{INAF Osservatorio Astrofisico di Arcetri, Largo Enrico Fermi 5, 50125 Firenze, Italy}
\affiliation{Dipartimento di Fisica e Astronomia, Universit\`a  degli Studi di Firenze, Via Sansone 1, 50019 Sesto Fiorentino (FI), Italy}
\affiliation{Istituto Nazionale di Fisica Nucleare, Sezione di Firenze, Via Sansone 1, 50019 Sesto Fiorentino (FI), Italy}
\email{niccolo.bucciantini@inaf.it}

\author[orcid=0000-0002-3382-9558]{B. M. Gaensler}
\affiliation{Department of Astronomy and Astrophysics, University of California, Santa Cruz, 1156 High Street, Santa Cruz, CA 95064, USA}
\affiliation{Dunlap Institute for Astronomy and Astrophysics, University of Toronto, 50 St. George Street, Toronto, ON, M5S 3H4, Canada}
\affiliation{David A. Dunlap Department of Astronomy and Astrophysics, University of Toronto, 50 St. George Street, Toronto, ON, M5S 3H4, Canada}

\email{gaensler@ucsc.edu}

\author[orcid=0000-0002-9322-9319]{Zhen Yan}
\affiliation{Shanghai Astronomical Observatory, Chinese Academy of Sciences, 80 Nandan Road, Shanghai 200030, China}
\email{yanzhen@shao.ac.cn}

\author[orcid=0000-0003-1526-7587]{David J. Wilner}
\affiliation{Center for Astrophysics \textbar{} Harvard \& Smithsonian, 60 Garden Street, Cambridge, MA 02138, USA}
\email{dwilner@cfa.harvard.edu}

\author[orcid=0000-0002-1389-765X]{Douglas C.-J. Bock}
\affiliation{CSIRO Space and Astronomy, PO Box 76, Epping, NSW 1710, Australia}
\email{Douglas.Bock@csiro.au}

\author[orcid=0000-0003-2641-9240]{Hua-bai Li}
\affiliation{Department of Physics, The Chinese University of Hong Kong, Shatin, New Territories, Hong Kong}
\email{hbli@phy.cuhk.edu.hk}

\shortauthors{Shi et al.}
\shorttitle{Low-energy spectral break of the Mouse PWN}

\begin{comment}
\author{Y.-Y.~Chan}\affiliation{Department of Physics, The University of Hong Kong, Pokfulam Road, Hong Kong}
\email{xx}
\author{Yifan Sun}\affiliation{Department of Physics, The University of Hong Kong, Pokfulam Road, Hong Kong}
\affiliation{Hong Kong Institute for Astronomy and Astrophysics, The University of Hong Kong, Pokfulam Road, Hong Kong}
\email{xx}
\end{comment}
\begin{abstract}

Bow-shock pulsar wind nebulae (PWNe) are synchrotron sources formed when the outflow of supersonic pulsars is confined by the surrounding interstellar medium. These sources are dominated by freshly injected particles, thus providing a unique laboratory for studying particle acceleration in relativistic outflows. The Mouse is a prototypical bow-shock PWN and is bright in the radio and X-ray bands, enabling detailed multi-wavelength spectral modeling. 
With data from 17 telescopes, the Mouse is detected from 118\,MHz to 353\,GHz, covering over 3 decades in frequency, and in X-rays from 0.5 to 30\,keV. A clear spectral break is identified at $\sim3.7$\,GHz.
The radio spectrum exhibits a rising trend at low frequency, peaks at the break, and then declines. We find a change in spectral index $\Delta\alpha=0.65\pm0.05$ across the break. We attribute this to an intrinsic cutoff or steepening in the injected particle distribution. This could indicate the characteristic energy of leptons leaving the pulsar magnetosphere, or energy dissipation after the particles cross the termination shock.

\end{abstract}
\keywords{pulsars: individual\ (PSR~J1747$-$2958) -- stars: neutron -- particle acceleration}

\section{Introduction}
\label{sec:intro}
Pulsar wind nebulae (PWNe) are broadband synchrotron sources powered by relativistic pulsar winds. When the wind is terminated by the surrounding medium, particles are accelerated to ultra-relativistic energies, producing emission from radio to $\gamma$-rays. Despite their efficiency as cosmic ray accelerators, the underlying acceleration mechanism in relativistic shocks remains theoretically challenging. Diffusive shock acceleration (DSA) has long been regarded as a standard mechanism for producing non-thermal power-law (PL) spectra as observed in astrophysical shocks \citep{bel78,bo78,pea81,hil84}. In highly magnetized relativistic shocks, classical DSA has traditionally been considered inefficient. Shock-driven magnetic reconnection has therefore been proposed as an alternative acceleration channel \citep{lk01,lyu03b,lgk+21}. Recent studies adopt a more realistic large-scale magnetic-field geometry in the striped-wind region. They find that DSA can operate efficiently in the equatorial region of the termination shock \citep{gk18,cg20}. DSA may therefore play a larger role than previously expected. Particle-in-cell (PIC) simulations of relativistic pair plasmas often predict particle spectra consisting of a quasi-thermal (Maxwellian-like) component connected to a non-thermal PL tail \citep[hereafter Maxwellian-PL;][]{spi08,2009ApJ...698.1523S}, implying a characteristic injection energy associated with the transition from thermal to non-thermal particles.

Radio-emitting particles have long synchrotron cooling times, such that they preserve the injected particle distribution.
In principle, the low-frequency spectrum is therefore a direct probe of the injection physics. In practice, however, the low-frequency spectral features seen in most PWNe are commonly attributed to absorption and to interactions with supernova remnant (SNR) ejecta, rather than to the intrinsic injection physics. Direct observational evidence for a low-energy steepening in the injected spectrum is therefore rare.

Bow-shock PWNe offer a unique opportunity to overcome this limitation. Because the pulsar has escaped its parent SNR and the nebula is approximately in steady state, the particle population is dominated by freshly injected pairs advected downstream of the shock. In such systems, the broadband synchrotron spectrum can more directly probe the current injection distribution.

Here we present multi-wavelength observations of the Mouse PWN, the prototype of bow-shock PWNe, combining data from 17 telescopes. We identify a spectral break at $\sim 3.7$\,GHz with a turnover $\Delta \alpha \approx 0.65$. More interestingly, the spectrum below the break is not merely flat, but rising. We demonstrate that this feature cannot be explained by synchrotron cooling, self-absorption, or free-free absorption. Instead, it requires a steepening in the injected particle spectrum at GeV energies, providing direct evidence for a low-energy injection threshold in a steady-state pulsar wind. In Section~\ref{sec:observation}, we give an overview of the Mouse and describe multi-wavelength observations used in this study.
In Section~\ref{sec:modeling}, we build a simple one-zone emission model to test different injected particle distributions.
The results are discussed in Section~\ref{sec:discuss} and summarized in Section~\ref{sec:Summary}.

% \section{Observations of the Mouse} 
\section{Multi-wavelength Spectrum of the Mouse}
\label{sec:observation}
\subsection{Previous studies}
The Mouse (G359.23$-$0.82) is a bright bow-shock PWN powered by the energetic pulsar J1747$-$2958.
The pulsar shows radio and $\gamma$-ray pulsations and has a high spin-down luminosity of $\dot E=2.5\times10^{36}$\ergs.
Its associated PWN was first discovered in radio with the Very Large Array \citep[VLA;][]{yb87} then in X-rays with \emph{ROSAT} and \cxo\ \citep{pk95,gvc+04}.
In both bands, the pulsar is embedded in a bright, compact nebula followed by a long tail.
The tail extends over 15\arcmin\ in radio but only $\sim45\arcsec$ in soft X-rays,
due to fast synchrotron burn-off of the high-energy particles.
The overall structure of the PWN resembles a classical bow shock and the pulsar standoff distance suggests highly supersonic motion \citep{gvc+04,kkp+18}.
The distance to the Mouse is estimated to be $\sim5$ kpc \citep{gvc+04}, for which the measured proper motion corresponds to a projected velocity of $\sim300$\kms \citep{hgc+09}. Throughout this work, we adopt a three-dimensional pulsar velocity of $v_{\rm psr}\approx600$\kms \citep{gvc+04}, since the bow-shock ram pressure and post-shock temperature are determined by the space velocity rather than its projected component.
A recent study with deep \cxo\ X-ray observations suggests an equipartition $B_{\rm pwn}$ of $\sim$250$\mu$G and a bulk flow speed of $\sim$4400\kms\ in the PWN \citep{kkp+18}. Our conclusion is not sensitive to these results.

\subsection{Data used in this work}
In this study, we make use of all available observations of the Mouse to compile a comprehensive spectral energy distribution (SED) using data from 17 telescopes.
The result is then employed to test different particle acceleration theories, using a physical model that accounts for the effects of particle acceleration and cooling. The data used in this work are summarized in Table~\ref{tab:data}. These include low-frequency radio, centimeter, millimeter/submillimeter, infrared, and X-ray band observations obtained from both archival surveys and new measurements:

\noindent $\bullet$  In the low-frequency radio band, we used archival data from the Murchison Widefield Array \citep[MWA;][]{2013PASA...30....7T}, and the Giant Metrewave Radio Telescope (GMRT; R.~Basu, private communication).
The MWA data are archival observations from the GLEAM survey \citep{2015PASA...32...25W,2017MNRAS.464.1146H}.

\noindent $\bullet$  In the centimeter band, the Mouse was observed with the Very Large Array \citep[VLA;][]{yb87,yb89,2000AJ....119..207L}, the Molonglo Observatory Synthesis Telescope (MOST; D.~Campbell-Wilson, private communication), the MeerKAT Galactic Centre Survey \citep{2022ApJ...925..165H}, and the Australian Square Kilometer Array Pathfinder \citep[ASKAP;][]{ASKAP1368}. We also incorporated two Australia Telescope Compact Array (ATCA) datasets at 1.38 and 94.5\,GHz (PI: Bock), along with new ATCA observations at 2.1, 5.5, 9.0, 16.7, and 21.2\,GHz obtained in 2023 (PI: Ng, see Appendix~\ref{Appendix:ATCA}). Further radio data include the Tian Ma Radio Telescope at 6.5, 16, 24, 43\,GHz (TMRT; PI: Yan).

\noindent $\bullet$  In the millimeter/submillimeter band, we made use of observations from the Berkeley-Illinois-Maryland Association (BIMA; PI: Bock) array, and the Atacama Cosmology Telescope (ACT) Data Release 6 (DR6) 2017--2022 night-only coadded maps at 98, 150, and 220\,GHz \citep{2025arXiv250314454C}.  The source was newly observed with the band~1, 3 and 4 of the Atacama Large Millimeter/submillimeter Array (ALMA) at 37, 39, 41, 43, 90.5, 92.5, 102.5, 104.5, 138, 140, 150, 152\,GHz (project 2025.1.01325.S; PI: Shi; see Appendix~\ref{Appendix:ALMA}), and an archival ALMA observation at 100\,GHz (project 2019.1.01187.S; PI: Klingler). We also made use of data from the Submillimeter Array (SMA; PI: Wilner).  We present new observations from the Institute for Radio Astronomy in the Millimetre Range (IRAM) 30m telescope, using the NIKA2 camera at 2 mm and 1.1 mm (PI: Shi; see Appendix~\ref{Appendix:IRAM}), along with the data from the James Clerk Maxwell Telescope (JCMT) at 850 $\mu$m and 450 $\mu$m (PI: Ng; see Appendix~\ref{Appendix:JCMT}).

\noindent $\bullet$  In the infrared band, we used archival \emph{Spitzer} data: images at 3.6, 4.5, 5.8, and 8.0\,$\mu$m taken with the Infrared Array Camera (IRAC), and 24 and 70\,$\mu$m observations from the Multiband Imaging Photometer for \emph{Spitzer} Galactic plane survey \citep[MIPSGAL;][]{2009PASP..121...76C, gh15}. More details can be found in Appendix~\ref{Appendix:Spitzer}.

\noindent $\bullet$  In the X-ray band, we included data from archival \emph{Chandra} observations \citep{kkp+18} and archival \emph{NuSTAR} data (ObsID: 30401036002; PI: Li). The \cxo\ observations analyzed in this work are available as a \cxo\ Data Collection (DOI:\dataset[10.25574/cdc.694]{https://doi.org/10.25574/cdc.694}). Further details are provided in Appendix~\ref{Appendix:Xray}.

\noindent $\bullet$ In the $\gamma$-ray band, the pulsar is detected in the Third \emph{Fermi}-LAT Catalog of Gamma-Ray Pulsars \citep{2023ApJ...958..191S}, with the emission attributed entirely to its pulsed emission rather than the PWN. We therefore do not include the $\gamma$-ray flux in our modeling of the PWN emission.

\begin{longtable*}[th!]{ccccc}
\caption{Flux Density Measurements of the Mouse\label{tab:data}}\\
\toprule
\multirow{2}{*}{Telescope} & \multirow{2}{*}{$\nu$/$\lambda$/Energy} & \multirow{2}{*}{Obs.\ Date} & \multirow{2}{*}{Array Config.}  & \multirow{2}{*}{$F_\nu$ (Jy)}\\
& & & & \\
\midrule
\endfirsthead

\multicolumn{5}{c}%
{{\bfseries \tablename\ \thetable{} Flux Density Measurements of the Mouse}} \\
\toprule
\multirow{2}{*}{Telescope} & \multirow{2}{*}{$\nu$/$\lambda$/Energy} & \multirow{2}{*}{Obs.\ Date} & \multirow{2}{*}{Array Config.}  & \multirow{2}{*}{$F_\nu$ (Jy)} \\
& & & &  \\
\midrule
\endhead

\bottomrule
\multicolumn{5}{r}{{Continued on next page}} \\
\endfoot

\bottomrule
\endlastfoot

MWA & 118\,MHz & 2014 Jun 13 & \nodata & $1.20\pm0.70$ \\
MWA & 150\,MHz & 2014 Jun 13 & \nodata & $1.50\pm0.50$ \\
GMRT & 153\,MHz & 2016 Mar 15 & hybrid & $1.18\pm0.18$ \\
MWA & 200\,MHz & 2014 Jun 13  & \nodata & $2.20\pm0.30$ \\
GMRT & 325\,MHz & 2016 Nov 11 & hybrid & $>0.15$ \\
VLA & 330\,MHz & 1986 Sep 15 & B, CnB & $1.43\pm0.01$ \\
GMRT & 610\,MHz & 2015 Aug 14 & hybrid  & $>0.29$ \\
MOST & 843\,MHz & 1999--2006  &\nodata & $1.76\pm0.11$ \\
GMRT & 1.28\,GHz & 2016 Dec 10 & hybrid & $1.49\pm0.22$ \\
MeerKAT & 1.28\,GHz & 2018 Jul 01 & \nodata & $1.81\pm0.14$ \\
ASKAP & 1.36\,GHz & 2021 Nov 20 & \nodata & $1.60\pm 0.25$ \\
ATCA & 1.38\,GHz & 2004 Mar 21 & 1.5A & $1.15\pm 0.27$ \\
VLA  & 1.47\,GHz  & 1987 Feb 20 & DnC & $1.69\pm0.24$ \\
ATCA & 1.84\,GHz & 2023 Apr 01 & 750C & $1.57\pm0.37$ \\
ATCA & 2.36\,GHz & 2023 Apr 01 & 750C & $1.96\pm0.32$ \\
ATCA & 2.87\,GHz & 2023 Apr 01 & 750C & $1.96\pm0.01$ \\
VLA & 4.86\,GHz  & 1987 Feb 22 & DnC & $1.54\pm0.05$ \\
ATCA & 4.98\,GHz  & 2023 May 13  & EW352 & $1.78\pm0.26$ \\
ATCA & 6.01\,GHz  & 2023 May 13 & EW352 &  $1.61\pm0.06$ \\
TMRT & 6.50\,GHz & 2018 Nov 11  &\nodata & $1.05\pm0.05$ \\	
VLA & 8.46\,GHz  & 1999 Oct 08 & BnA  & $>0.32$ \\
ATCA & 8.49\,GHz  & 2023 May 13 & EW352 &  $1.39\pm0.02$ \\
ATCA & 9.51\,GHz  & 2023 May 13 & EW352  & $1.36\pm0.05$ \\
VLA & 14.9\,GHz & 1988 Feb 25 & CnB  & $>0.31$ \\
TMRT & 16.0\,GHz & 2018 Sep 30 & \nodata & $0.55\pm 0.03$ \\ 
ATCA & 16.7\,GHz  & 2023 Sep 10 & H168 & $>0.78$ \\
ATCA & 21.2\,GHz  & 2023 Sep 10 & H168 & $>0.62$ \\
TMRT & 24.0\,GHz & 2019 Jan 18 & \nodata & $0.66\pm0.18$ \\
%ALMA & 37, 39, 41, 43\,GHz & 2025 Oct 24 & 7\,m & $> 0.27$ \\
ALMA & 37.0\,GHz & 2025 Oct 24 & 7\,m & $> 0.35$ \\
ALMA & 39.0\,GHz & 2025 Oct 24 & 7\,m & $> 0.34$ \\
ALMA & 41.0\,GHz & 2025 Oct 24 & 7\,m & $> 0.30$ \\
ALMA & 43.0\,GHz & 2025 Oct 24 & 7\,m & $> 0.27$ \\
TMRT & 43.0\,GHz & 2019 Jan 17 & \nodata & $0.70\pm0.38$ \\
BIMA & 86.9\,GHz &2004 Apr 19  & \nodata  &  $>0.1$ \\
%ALMA & 90, 92, 102, 104\,GHz & 2026 Jun 22 & 7\,m & $>0.12$ \\
ALMA & 90.5\,GHz & 2026 Jun 22 & 7\,m & $>0.17$ \\
ALMA & 92.5\,GHz & 2026 Jun 22 & 7\,m & $>0.13$ \\
ATCA & 94.5\,GHz & 2005 Jul 11 & H75 &  $>0.03$ \\
ACT & 98.0\,GHz & 2017--2022 & \nodata & $0.51\pm0.06$ \\
ALMA & 100\,GHz & 2019 Dec 12 & C43-1 &  $>0.12$ \\
ALMA & 102.5\,GHz & 2026 Jun 22 & 7\,m & $>0.13$ \\
ALMA & 104.5\,GHz & 2026 Jun 22 & 7\,m & $>0.12$ \\
%ALMA & 138, 140, 150, 152\,GHz & 2026 Apr. 20 & 7\,m & $>0.09$ \\
ALMA & 138\,GHz & 2026 Apr 20 & 7\,m & $>0.12$ \\
ALMA & 140\,GHz & 2026 Apr 20 & 7\,m & $>0.11$ \\
ACT & 150\,GHz & 2017-2022 & \nodata & $0.42\pm0.16$ \\
ALMA & 150\,GHz & 2026 Apr 20 & 7\,m & $>0.09$ \\
ALMA & 152\,GHz & 2026 Apr 20 & 7\,m & $>0.09$ \\
IRAM 30m & 160\,GHz & 2025 Oct 24 & \nodata & $0.28\pm0.05$ \\
ACT & 220\,GHz & 2017-2022 & \nodata & $<0.89$ \\
SMA & 224\,GHz & 2004 May 24  &\nodata & $>0.018$ \\
IRAM 30m & 260\,GHz & 2025 Oct 24 & \nodata & $0.21\pm0.01$ \\
JCMT & 850\,$\mu$m & 2025 Jul 12--14 & \nodata & $0.15\pm{0.02}$ \\
JCMT & 450\,$\mu$m & 2025 Jul 12--14 & \nodata & $<0.50$ \\
\emph{Spitzer} & 71.4$\mu$m & 2005 Apr 8 & \nodata & $<0.90$ \\
\emph{Spitzer} & 23.7$\mu$m & 2005 Apr 8 & \nodata &  $<0.03$ \\
\emph{Spitzer} & 8.0$\mu$m & 2005 Sep 15 & \nodata &  $<0.18$ \\
\emph{Spitzer} & 5.8$\mu$m & 2005 Sep 15 & \nodata &  $<0.6$ \\
\emph{Spitzer} & 4.5$\mu$m & 2005 Sep 15 & \nodata &  $<0.016$ \\
\emph{Spitzer} & 3.6$\mu$m & 2005 Sep 15 & \nodata &  $<0.016$ \\
\emph{Chandra} & 0.5--8\,keV & 2002, 2013, 2014 & \nodata &  \multicolumn{1}{l}{see Appendix~\ref{Appendix:Xray}} \\
\emph{NuSTAR} & 3--79\,keV & 2018 Jul 29 & \nodata & \multicolumn{1}{l}{see Appendix~\ref{Appendix:Xray}} \\

\end{longtable*}

\iffalse
\begin{figure}[tp!]
	\centering
	\includegraphics[width=0.49\textwidth]{Figures/ASKAP.png}
	\caption{ASKAP image of the Mouse at 1.36\,GHz. The solid ellipse and dashed circle show the region used for flux density measurement of source and background subtraction. The color scale is in units of Jy/beam.}
	\label{fig:ASKAP}
\end{figure}
\fi

\begin{figure}[tp!]
	\centering
	\includegraphics[width=0.47\textwidth]{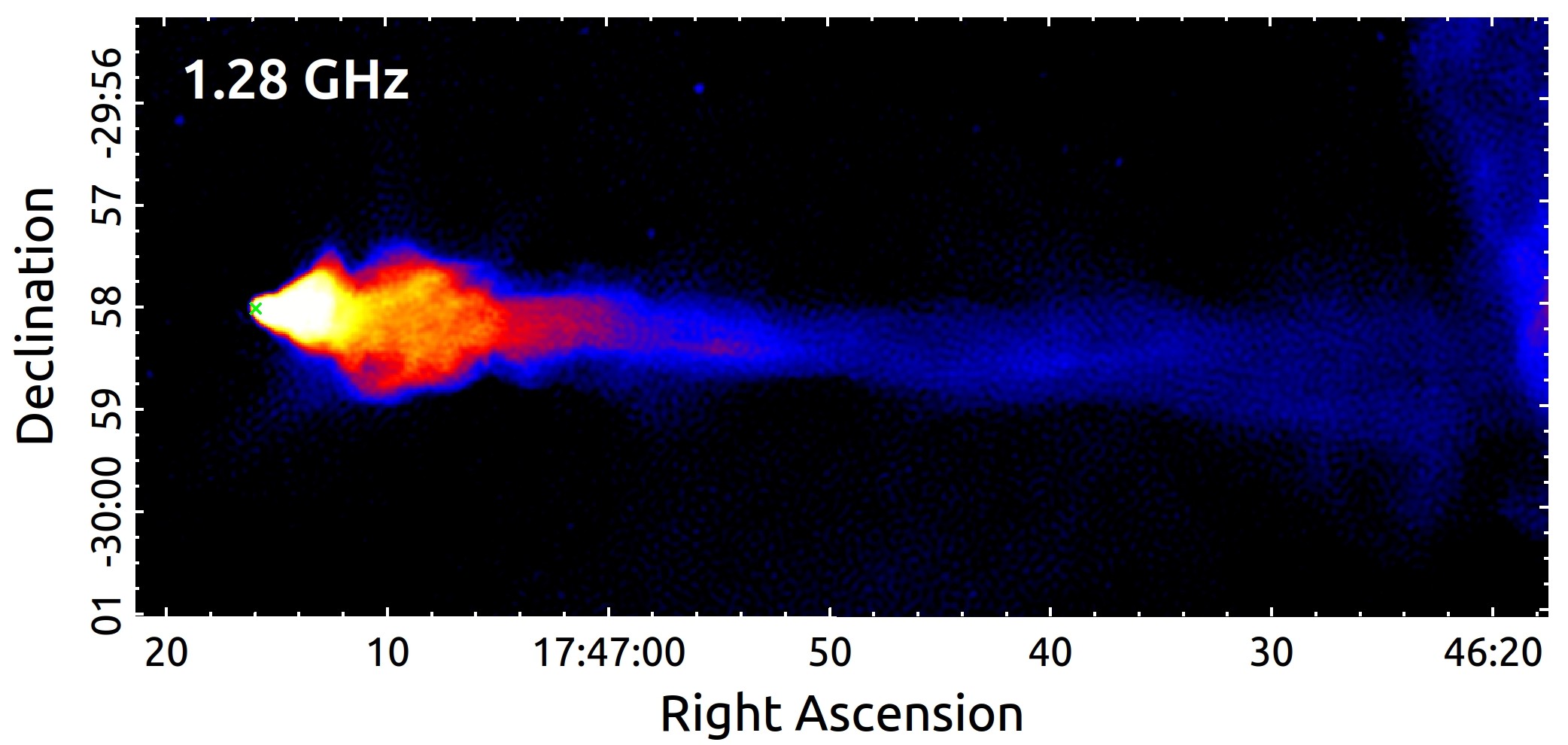}
	\caption{MeerKAT image of the Mouse at 1.28\,GHz obtained from MeerKAT Galactic
Centre Survey \citep{2022ApJ...925..165H}. The color scale is in units of mJy/beam. The green cross denotes the position of the pulsar.} %The solid ellipse and dashed circle show the region used for flux density measurement of source and background subtraction. }
	\label{fig:meerkat}
\end{figure}

We measured the flux density by defining the same region enclosing the nebula across all images and selecting several elliptical background regions to estimate and subtract the background. %using {\sc ds9} \citep{2003ASPC..295..489J}. 
Background variations dominate the uncertainties in the measured results. The measurements from the ATCA 94.5\,GHz, SMA, ALMA (Bands~1, 3, and 4), and BIMA datasets suffer from flux deficits arising from the inherent short-spacing limitation, a direct consequence of incomplete sampling of the $uv$-plane. This restricted $uv$-coverage effectively filters out spatially smooth, extended emission components. So the total integrated flux density measurements of these observations are taken as lower limits. The ACT 220\,GHz and \emph{Spitzer} images give upper limits only due to non-detection or extraneous background radiation.

\begin{figure*}[tp!]
	\centering
	\includegraphics[width=0.85\textwidth]{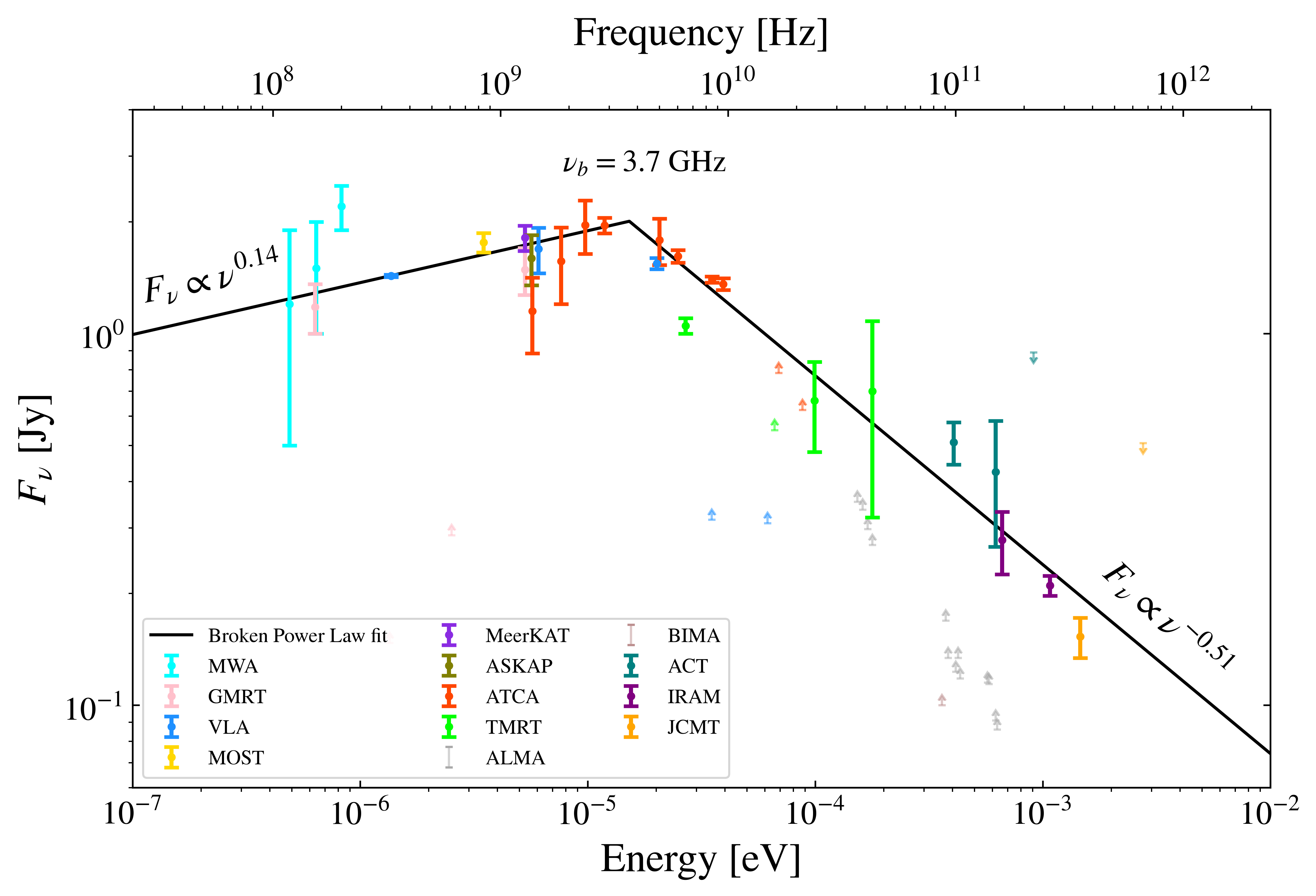}
	\caption{Radio-to-submillimeter spectrum of the Mouse with the best-fit broken PL model (black solid line). The data points are measurements from MWA (118, 150, 200\,MHz), GMRT (153, 325, 610\,MHz, 1.28\,GHz), VLA (330\,MHz, 1.5, 4.9, 8.4, and 14.9\,GHz), MOST (843\,MHz), MeerKAT (1.3\,GHz), ASKAP (1.36\,GHz), ATCA(1.38, 4.98, 6.01, 8.48, 9.51, 16.7, 21.2, 94.5\,GHz), TMRT (6.5, 16, 24, 43\,GHz), ALMA (37, 39, 41, 43, 90, 92, 100, 102, 104, 138, 140, 150, 152\,GHz), BIMA (86.9\,GHz), ACT (98, 150, 220\,GHz), IRAM 30m (160, 260\,GHz), and JCMT (353, 666\,GHz).\label{fig:spec}}
\end{figure*}

\subsection{Results \label{subsec:results}}
The flux density measurement results are listed in Table~\ref{tab:data} and are plotted in Figures~\ref{fig:spec} and~\ref{fig:SED}. Note that we did not subtract the
contribution from the pulsar, except in the X-ray band. We argue this has a negligible effect since the pulsar in the radio to submillimeter bands is
2--3 orders of magnitude fainter than the nebula \citep{2018MNRAS.475.1469B}.
The multi-wavelength spectrum in Figure~\ref{fig:spec} clearly shows a break at a few GHz.
Below the break, the spectrum is relatively flat and gradually rising. It then
abruptly steepens and then declines above the break.
%such that it is unclear if the spectrum shows any curvature.
We fit the spectrum with two simple models: a single PL,
\begin{equation}\label{eqn:photonsinglepl}
	F_\nu = A\nu^{-\alpha}~,
\end{equation}
and a broken PL,
\begin{equation}\label{eqn:photonbrokenpl}
	F_\nu = A\begin{cases}
\nu^{-\alpha_1}~ & \text{$\nu < \nu_b$} \\
\nu_b^{\alpha_2-\alpha_1}\nu^{-\alpha_2}~ & \text{$\nu > \nu_b$}
\end{cases}~,
\end{equation}
where $A$ is the normalization factor. The frequencies $\nu$ and $\nu_b$ are in units of GHz. The single PL is the special case $\alpha_1=\alpha_2$ of the broken PL. We fit both models to the radio-to-submillimeter measurements shown in Figure~\ref{fig:spec}.

We obtain the best-fit model parameters by the maximum likelihood estimation method. The log-likelihood of the model $M(E_i, \nu_i)$ is defined as
\begin{equation}
	\ln \mathcal{L}=-\sum_{i=1}^{N}\frac{\left|\log\big(M(E_i, \nu_i)\big)-\log(D_i)\right|^2}{2\sigma_{\log,i}^2}~,
\end{equation} where $D_i$ and $\sigma_{\log,i}$ are observational data and uncertainty in log-scale of energy $E_i$ or frequency $\nu_i$ over $N$ measurements. The broken PL describes the data far better than the single PL ($\ln\mathcal{L}=-57$ vs.\ $-568$). An $F$-test comparing the two nested fits gives $F(2,23)=104$, corresponding to $p=3\times10^{-12}$. The broken PL is also favored by $\Delta{\rm AIC}\approx1018$. The break is therefore highly significant.
The black solid line in Figure~\ref{fig:spec} shows the best-fit broken PL, with a spectral break $\nu_b = 3.7^{+0.7}_{-0.6}$\,GHz. Below this frequency, the spectral index is $\alpha_1 =-0.14\pm0.04$, indicating a rising spectrum. Above the break, the spectrum steepens to $\alpha_2 = 0.51\pm0.03$, corresponding to a spectral index change of $\Delta\alpha = \alpha_2-\alpha_1= 0.65\pm0.05$ across the transition. 

To investigate whether the break arises from spatially distinct emission regions of the Mouse (e.g., the head and the tail), we extracted the radio spectrum of the head region alone. The spectral break, previously identified in the full nebula spectrum, persists in the head-only spectral profile. This result implies that the spectral break is an intrinsic property, rather than a consequence of combining emission from different regions.
% \textbf{We find that the pulsar is at least two orders of magnitude fainter than the surrounding PWN, based on flux density measurements from \cite{2018MNRAS.475.1469B} and the ATNF catalog \citep{2005AJ....129.1993M}.
% This effectively rules out pulsar variability as the origin of the spectral break.}
In Section~\ref{subsec:breaknature}, we will demonstrate that the observed spectral break cannot be attributed to synchrotron cooling, synchrotron self-absorption, or free-free absorption.

\section{Modeling the multiwavelength SED}
\label{sec:modeling}
\subsection{Method and parameters}

As introduced in Section~\ref{sec:intro}, bow-shock PWNe can be considered as in a
steady state. Here, we develop a one-zone steady model to investigate the
injected particle distribution. The Mouse's cometary morphology is depicted in
Figure~\ref{fig:meerkat}, and we simplify the model by treating the nebula as a cylinder. The maximum energy of the particles is taken as
$\sim3\,$PeV, estimated with the maximum electric potential given by the polar
cap of the pulsar
\begin{equation}
    E_{\rm max}=\Delta\Phi\approx\frac{BR_{\rm psr}^3\Omega^2}{c^2}~,
\end{equation}where $B$, $R_{\rm psr}$, and $\Omega$ are the surface magnetic field, radius, and angular velocity of the pulsar. 
Given that particles at this maximum energy emit radiation above the X-ray band, this choice exerts no influence on the modeling results. The modeling parameters are listed in Table~\ref{tab:modelpars}.

\begin{figure*}[tp!]
	\centering
    \includegraphics[width=0.99\textwidth]{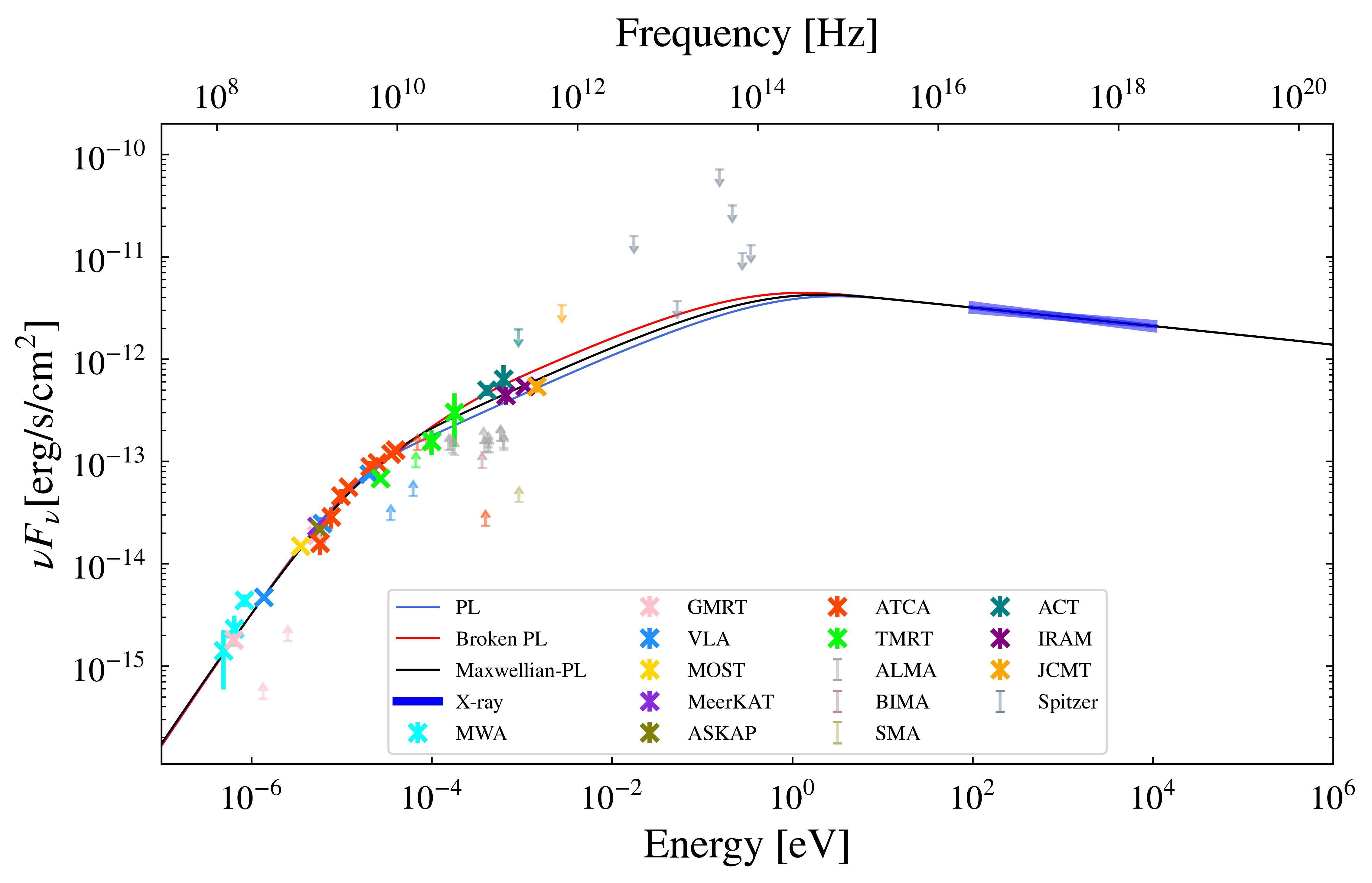}
\caption{Multi-wavelength SED of the Mouse, covering radio, submillimeter,
infrared, and X-ray measurements or limits. The best-fit single PL with low cut,
broken PL with low cut, and Maxwellian-PL injection models are plotted in
blue, red, and black solid lines, respectively. The X-ray \cxo\ results are shown in a blue envelope. }
\label{fig:SED}
\end{figure*}

\begin{table}[!ht]
\centering
\caption{Model Parameters\label{tab:modelpars}}
\hspace*{-1cm}
\begin{tabular}{ccc}
	\toprule
	\multicolumn{2}{c}{Modeling parameters} & References\\
	\midrule
		%$B_{\rm pwn]$ & $250\,\mu$G & \cite{kkp+18} \\
		$v_{\rm flow}$ & $4400$\kms & \cite{kkp+18} \\
		$E_{\rm max}$ & $\sim3\,$PeV & \nodata\\
		$d$ & $5\,$kpc & \citet{gvc+04} \\
        $v_{\rm psr}$ & $600$\kms & \citet{gvc+04} \\
		$l_{\rm tail}$ & $15\farcm0$ & \nodata\\
		%$w$ & $3\farcm7$ & \nodata\\
	\bottomrule
	\end{tabular}
\end{table}

We consider three distributions for the injected particles:
\begin{itemize}[leftmargin=0.8em, label=\scriptsize$\bullet$]
\item Single PL with a low-energy cutoff as suggested by conventional DSA theory:
\end{itemize}
\begin{equation}
\label{eqn:singlePL}
\frac{{\rm d}N}{{\rm d}E}=N_0
\left\{
\begin{array}{lr}
    0  & E<E_{\rm min},\\
    E^{-p}                        &E_{\rm min} < E < E_{\rm max},\\
    0                        &E > E_{\rm max}.
\end{array}
\right.
\end{equation}

\begin{itemize}[leftmargin=0.8em, label=\scriptsize$\bullet$]
\item Broken PL with a low-energy cutoff, as suggested by magnetic reconnection with DSA:
\end{itemize}
\begin{equation}
\label{eqn:brokenPL}
\frac{{\rm d}N}{{\rm d}E}=N_0
\left\{
\begin{array}{lr}
    0  & E<E_{\rm min},\\
    E^{-p_0}                       &E_{\rm min}<E < E_{\rm bre},\\
    E_{\rm bre}^{p-p_0}E^{-p}    &E_{\rm bre} < E < E_{\rm max},\\
    0                       &E > E_{\rm max}.
\end{array}
\right.
\end{equation}

\begin{itemize}[leftmargin=0.8em, label=\scriptsize$\bullet$]
\item Relativistic Maxwellian and a high-energy PL tail with exponential cutoff,
as suggested by PIC simulations of relativistic shocks 
\citep[Maxwellian-PL;][]{2009ApJ...698.1523S}: 

${\rm d}N/{\rm d}E=N_0\times$
\end{itemize}
\begin{equation}
\left\{
\begin{array}{lr}
    E\sqrt{E^2-1}\exp{(-\frac{E}{\Delta E_{\rm MB}})}  & E<E_{\rm min},\\
    CE^{-p}                       &E_{\rm min}<E < E_{\rm cut}, \\
    CE^{-p}\exp{(-\frac{(E-E_{\rm cut})}{\Delta E_{\rm cut}})} & E > E_{\rm cut}.
\end{array}
\right.
\label{eqn:maxwell}
\end{equation} where $C=E_{\rm min}^{p+1}\sqrt{E_{\rm min}^2-1}\exp{(-E_{\rm min}/\Delta E_{\rm MB})}$ and $E_{\rm cut}$, $\Delta E_{\rm cut}$ are taken to be same as $E_{\rm max}\approx3$\,PeV for consistency, which has no impact on the results. In the models described above, $N_0$ is in units of eV$^{-1}$; the energies $E$, $E_{\rm min}$, $E_{\rm bre}$, $E_{\rm cut}$, $\Delta E_{\rm cut}$ and $\Delta E_{\rm MB}$ are in units of eV. 

The characteristic synchrotron frequency of a particle with energy $E$ is
$\nu_{\rm cri} = 3e/(4\pi m_e^3c^5)BE^2\sin\theta$. The total synchrotron power radiated by the particle $P_{\rm syn}$ is given by the standard formula
\begin{equation}\label{P_syn}
    P_{\rm syn} = \frac{2e^4}{3m_e^4c^7}B_{\rm pwn}^2\sin^2\theta E^2
\end{equation} \citep{pac70}, where $e$ and $m_e$ are the electron charge and
mass, respectively, $c$ is the speed of light, $E$ is the particle energy,
$\theta$ is the pitch angle, and $\sin\theta$ is taken to be $\sqrt{2/3}$
assuming isotropic pitch angle distribution. We account for synchrotron cooling
by evolving the particle energies over time before calculating the
flux: the particle age $t_{\rm age}=l_{\rm tail}/v_{\rm flow}$ is divided into
100 even intervals in log scale, to form 100 particle groups with varying ages
($t_1, t_2,\ldots,t_{100}$) and energies ($E({t_1}),
E({t_2}),\ldots,E({t_{100}})$). As the particle cool-down rate is $P_{\rm
syn}$, i.e.\ 
\begin{eqnarray}\label{eqn:cool}
	\frac{{\rm d}E(t)}{{\rm d}t}=-P_{\rm syn}\propto -E^2,
\end{eqnarray}
the integral of the above equation gives
\begin{eqnarray}\label{eqn:ncool}
	\frac{1}{E({t_{n+1})}}-\frac{1}{E({t_{n})}} \propto t_{n+1}-t_{n}~,
\end{eqnarray} where $n=1,2,\ldots,99$. We iteratively compute the energy distribution $E({t_n})$ for each group.

The total synchrotron flux density $F_\nu$ is the sum of contributions from all 100 groups, including old and young particles, calculated individually with the \texttt{naima} Python package \citep{naima}. This calculation method results in $\Delta p=1$
between the cooled and uncooled particles, consistent with the standard value expected from synchrotron theory.

Due to the different sampling and uncertainty estimation methods between the radio-to-infrared and X-ray datasets, we adopted a two-step approach instead of a combined fit to avoid potential biases and maintain self-consistency. First, we use the X-ray spectrum results (see details in Appendix~\ref{Appendix:Xray}), as they provide tighter constraints on the high-energy spectrum. 
The X-ray fit yields a photon index of $\Gamma_X = 2.09\pm0.02$, corresponding to an underlying particle index of $p_X = 2\Gamma_X - 1 = 3.17$.
We then model the low-energy (radio-to-infrared) part of the spectral energy distribution (SED) using different
injected particle spectra (Eqs.~\ref{eqn:singlePL}, \ref{eqn:brokenPL}, and
\ref{eqn:maxwell}), with the normalization $N_0$ and particle index $p$
constrained by the X-ray results. Specifically, $N_0$ is scaled to match the
X-ray flux, and $p$ is fixed at the slow-cooling value $p = p_X - 1 = 2.17$ for
all three models. This approach minimizes biases from different data quality
while maintaining consistency across wavebands. Finally, we also fit the PWN
$B$-field, $B_{\rm pwn}$, which is a critical parameter for the synchrotron
radiation power and the cooling break frequency. Therefore, we only have two
free parameters  for the single PL fit $(E_{\rm min}, B_{\rm pwn})$, four for
the broken PL fit $(E_{\rm min}, B_{\rm pwn}, E_{\rm bre}, p_0)$, and three for the
Maxwellian-PL fit $( E_{\rm min}, B_{\rm pwn}, \Delta E_{\rm MB}$). 

\subsection{Modeling Results}

\begin{table}[t]
	\centering
	\caption{Best-fit parameters, log-likelihood ($\ln\mathcal{L}$) and AIC statistic of different particle distribution models. Both PL models include a low-energy cutoff.\label{tab:fitpars}}
	\footnotesize
	\setlength{\tabcolsep}{4.5pt}
	\begin{tabular}{lccc}
\toprule
 & Single PL & Broken PL & Maxwellian-PL\\
\midrule
$N_0$ (eV$^{-1}$) & $1.2\times10^{60}$ & $9.1\times10^{52}$ & $1.2\times10^{21}$\\
$E_{\rm min}$ (GeV) & 2.2 & 1.4 & 4.0\\
$\Delta E_{\rm MB}$ (GeV) & \nodata & \nodata & 1.5\\
$p_0$ & \nodata & 1.47 & \nodata\\
$p$ & 2.17 & 2.17 & 2.17\\
$E_{\rm bre}$ (GeV) & \nodata & 9.7 & \nodata\\
$B_{\rm pwn}$ ($\mu$G) & 41 & 55 & 46\\
\addlinespace
$\ln\mathcal{L}$ & $-59.3$ & $-58.9$ & $-44.7$\\
AIC & 122.6 & 125.8 & 95.4\\
\bottomrule
\end{tabular}
\end{table}

\begin{figure*}[tp!]
	\centering    \includegraphics[width=0.95\textwidth]{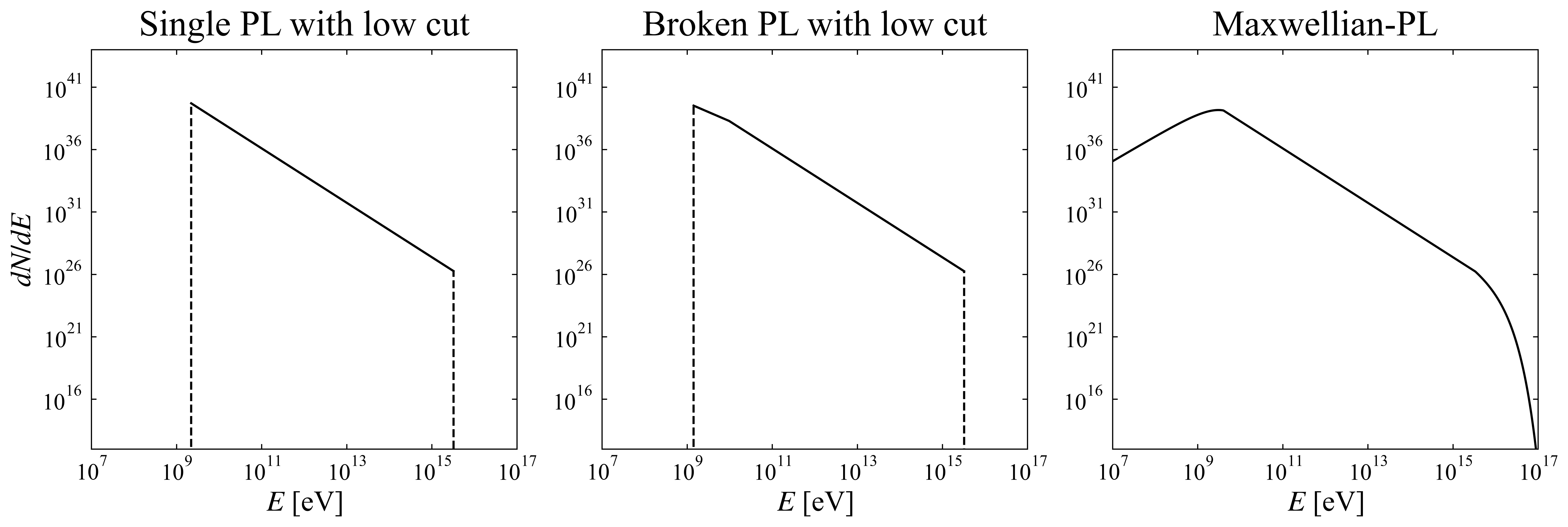}
\caption{Best-fit particle spectrum of single PL, broken PL and Maxwellian-PL models. The best-fit parameters are listed in Table~\ref{tab:fitpars}.}
\label{fig:partspec}
\end{figure*}

The best-fit models are plotted in Figure~\ref{fig:SED} and their parameters are listed in Table~\ref{tab:fitpars}.
Figure~\ref{fig:partspec} shows the best-fit injected particle distributions.
The low-energy turnover ($E_{\rm min}$ around 2 and 1\,GeV in the single and broken
PL, and $\Delta E_{\rm MB}=1.5$\,GeV in the Maxwellian-PL model) well reproduces the observed spectral break 
at 3.7\,GHz. The pulsar wind emission from particles of different ages is superposed along the tail. The resulting spectrum bends over gradually between $10^{13}$ and $10^{16}\,\rm Hz$, i.e., between the infrared and the soft X-ray bands. The inferred magnetic fields of three models range from 40 to 60\,$\mu\rm G$. In the broken PL model, an additional low-energy PL component ($p_0 = 1.47$) is presented from 1.4 to 10\,GeV.

An $F$-test was performed to compare the goodness-of-fit of the single PL model (RSS=119, \text{df}=25) and the broken PL model (RSS=118, \text{df}=23), where RSS denotes the residual sum of squares and \text{df} denotes the degrees of freedom. The test yields a statistic of $F(2,23)=0.078$ and a corresponding $p$-value of $0.93>0.05$, indicating that the additional broken PL component does not significantly improve the fitting.
Furthermore, we calculated the Akaike Information Criterion (AIC $=2k-2\ln\mathcal{L}$) for the three models considered, and the values are listed in Table~\ref{tab:fitpars}. The Maxwellian-PL model is preferred based on the AIC statistic.

\section{Discussion  \label{sec:discuss}}
\subsection{Nature of the Spectral Break \label{subsec:breaknature}}
Using archival and new observations from various telescopes, we measure the SED
of the Mouse with good frequency coverage from radio to the submillimeter bands. We
found clear evidence of a spectral break at $\sim$3.7\,GHz, confirming the result
from the previous study of \cite{kkp+18}. We argue below that this cannot be
due to synchrotron cooling, self-absorption, or free-free absorption.

\noindent $\bullet$ Synchrotron cooling can produce a spectral break at frequency

\begin{equation}
\nu_c({\rm GHz})=1.187 B_{\rm pwn}^{-3}({\rm G})t_{\rm age}^{-2}({\rm yr})
\end{equation}

\citep{2006ApJ...638..225K}. Here, $t_{\rm age}=l_{\rm tail}/v_{\rm flow}=4.8$\,kyr, where $l_{\rm tail}\approx15'$ is the length of the radio tail measured in this work and $v_{\rm flow}=4400$\kms\ \citep{kkp+18}. In our fitted models, the cooling break falls at $h\nu_c\approx2$--$3$\,eV. This is comparable to the estimate $h\nu_c\sim10$\,eV of \citet{kkp+18}. It lies about five orders of magnitude above the observed break.

The adopted flow speed does not affect this conclusion. \citet{kkp+18} calibrated $v_{\rm flow}$ over the X-ray-bright part of the tail, which is much shorter than the radio tail. We therefore refit all three models with $v_{\rm flow}$ fixed at values from $v_{\rm psr}\approx600$\kms\ \citep{gvc+04}, below which no trailing tail would form, up to $44{,}000$\kms. Over this range the fits follow $B_{\rm pwn}\propto v_{\rm flow}^{2/3}$ and $E_{\rm min}\propto v_{\rm flow}^{-1/3}$. The predicted spectrum and the fit quality are essentially unchanged, although $B_{\rm pwn}$ varies from 12 to $220\,\mu$G. The cooling break is a fitted quantity rather than an assumed one. Since $\nu_c\propto B_{\rm pwn}^{-3}t_{\rm age}^{-2}$ and $t_{\rm age}\propto v_{\rm flow}^{-1}$, the two dependences on $v_{\rm flow}$ cancel. The cooling break therefore stays within $5$--$7\times10^{14}$\,Hz across all models and flow speeds considered.

\noindent $\bullet$ Synchrotron self-absorption produces a spectral turnover at

\begin{equation}
\nu_{\rm SA}=\nu_L\left[\frac{\pi\sqrt{\pi}}{4}\frac{eRK}{B_{\rm pwn}}f_{\alpha}(p)\right]^{\frac{2}{p+4}}\label{eqn:nusa}
\end{equation}

\citep[Eq.~4.56 of][]{2013LNP...873.....G}, where $f_\alpha(p)$ is a dimensionless function of the particle index, $R$ is the size of the emitting region, $K$ is the normalization of the particle distribution $N(\gamma)=K\gamma^{-p}$, and $\nu_L\equiv eB_{\rm pwn}/2\pi m_ec$ is the Larmor frequency. We identify $R$ with the contact-discontinuity standoff radius, $R_{\rm CD}\approx0.024$\,pc \citep{gvc+04}. This gives $\nu_{\rm SA}\approx10$\,MHz, an order of magnitude below our lowest-frequency measurement and more than two orders of magnitude below the observed break. At fixed flux, $K\propto R^{-3}$. Equation~\ref{eqn:nusa} then gives $\nu_{\rm SA}\propto R^{-4/(p+4)}$. Our choice of $R$ is the smallest plausible scale for the emitting region. The radio emission instead fills the whole $15'$ tail. Thus, this choice gives an upper bound on $\nu_{\rm SA}$.

This estimate depends on the assumed emitting volume and source distance. We therefore use a complementary, distance-independent criterion. Synchrotron self-absorption becomes important when the brightness temperature approaches the effective temperature of the radiating electrons,

\begin{equation}
T_{\rm b}=\frac{c^2}{2k}\frac{F_\nu}{\nu^2\Omega}\lesssim T_{\rm eff}\simeq\frac{\gamma m_ec^2}{3k}~,
\label{eqn:tb}
\end{equation}

\citep[e.g.,][]{2013LNP...873.....G}. Here, $\Omega$ is the solid angle subtended by the source, and $3kT_{\rm eff}$ is the mean energy per electron. The Lorentz factor of the electrons radiating at frequency $\nu$ is $\gamma\simeq(2\nu/3\nu_L)^{1/2}$. This follows by inverting the characteristic synchrotron frequency $\nu_{\rm cri}$ defined in Section~\ref{sec:modeling}. Thus, $T_{\rm b}$ is set by the observed flux density and angular size. $T_{\rm eff}$ is set by $B_{\rm pwn}$. Neither depends on the source distance or emitting volume.

If the break at 3.7\,GHz were caused by self-absorption, the responsible electrons would have $\gamma\approx4700$--$3800$ for $B_{\rm pwn}=40$--$60\,\mu$G. The source would then have to reach $T_{\rm b}\sim8\times10^{12}$\,K. The flux density measured at the break ($\sim1.8$\,Jy) is spread over the $\sim15'\times2'$ tail resolved in our MeerKAT and ASKAP maps. It gives $T_{\rm b}\approx2$\,K, twelve orders of magnitude below the required value. The entire flux would have to come from a region $\lesssim2\times10^{-4}$\arcsec\ across to reach this value. The magnetic field affects the threshold only through $T_{\rm eff}\propto B_{\rm pwn}^{-1/2}$. Thus, even an order-of-magnitude error in $B_{\rm pwn}$ changes the threshold by only a factor of three. $T_{\rm eff}/T_{\rm b}$ decreases towards lower frequencies, following $T_{\rm eff}/T_{\rm b}\propto\nu^{\alpha+5/2}$. However, it remains $\sim10^9$ at 118\,MHz, our lowest-frequency measurement.

The spectral shape gives the same conclusion. Below the turnover, a self-absorbed synchrotron source is optically thick and has $F_\nu\propto\nu^{5/2}$. A turnover at 3.7\,GHz would therefore give only $\sim0.4$\,mJy at 118\,MHz. This is about three thousand times below the observed flux density. Instead, we measure a smooth rise, $F_\nu\propto\nu^{0.14\pm0.04}$ (Section~\ref{subsec:results}). The MWA points and the break span a factor of $\sim30$ in frequency. The spectrum shows no rollover across this range. A flat spectrum can also be produced by a superposition of self-absorbed components with different turnover frequencies, as in compact extragalactic jets. Each such component must be self-absorbed at GHz frequencies. That requires it to be compact and bright. The Mouse contains no such components. The nebula is resolved at all frequencies of interest. The only compact source in the field is the pulsar, and it is at least two orders of magnitude fainter than the PWN \citep{2018MNRAS.475.1469B,2005AJ....129.1993M}.

\noindent $\bullet$ Free-free absorption by the shocked ISM can also suppress the low-frequency emission. We estimate this effect following \citet{rl79} and \citet{2013LNP...873.....G}. The absorption coefficient is

\begin{equation}
\alpha_\nu^{\rm ff}=3.7\times10^8\frac{Z^2n_en_i}{T^{1/2}}\frac{1-e^{-h\nu/kT}}{\nu^3}\bar g_{\rm ff},
\end{equation}

where $Z\simeq1$ is the atomic number and $\bar g_{\rm ff}\simeq1$ is the Gaunt factor. For the unshocked gas we adopt the canonical interstellar value $n_0\approx1\,{\rm cm^{-3}}$. A strong adiabatic shock compresses this by a factor of four. The shocked region then has equal electron and ion densities, $n_e\simeq n_i\simeq4\,{\rm cm^{-3}}$. The post-shock temperature follows from the standard strong-shock jump conditions \citep[e.g.,][]{zr67,2012A&ARv..20...49V},

\begin{equation}
kT=\frac{3}{16}\mu_p m_pV_s^2\approx1.2\left(\frac{V_s}{1000\,{\rm km\,s^{-1}}}\right)^2\,{\rm keV},
\end{equation}

where $\mu_p\approx0.6$ is the mean molecular weight for solar abundances and $m_p$ is the proton mass. Taking $V_s$ as the pulsar velocity gives $T\approx5.0\times10^6$\,K.

The turnover occurs where $\tau_\nu\equiv\alpha_\nu^{\rm ff}L=1$. The thickness of the absorbing layer satisfies $L<R_{\rm CD}$. Adopting the maximum thickness gives $\nu_{\rm ff}\lesssim1.4$\,kHz. This is six orders of magnitude below the observed break. The turnover frequency scales as $\nu_{\rm ff}\propto(n_en_iL)^{1/2}T^{-3/4}$. Moving it to 3.7\,GHz would require $n_0\sim10^6\,{\rm cm^{-3}}$, far above the densities found even in dense molecular cloud cores. The spectral shape again gives an independent constraint. Below the turnover, free-free absorption produces an optically thick spectrum with $F_\nu\propto\nu^{2}$. This is much steeper than the rise we observe.

Based on the above estimates, we conclude that the observed spectral break is
unlikely due to cooling or absorption effects. Our modeling results in
Section~\ref{sec:modeling} shows that the break can be explained by a cutoff or
steepening of the underlying particle energy distribution at a few GeV.

% The rising spectrum ($F_{\nu}\propto \nu^{0.14}$) below the break strongly favors that the feature is caused by a low-energy steepening, rather than other physical mechanisms.

\subsection{Physical Interpretation of the Break}
\label{sec:physics}
We suggest that the drop in the number of low-energy particles could be due to
the acceleration process, either at the pulsar magnetosphere or at the wind
termination shock. In the first scenario, the cutoff at several GeV could
reflect the characteristic energies of particles leaving the pulsar. Global PIC
simulations demonstrate that most positrons are accelerated to $\gamma\sim10^3$
(i.e.\ $\sim$GeV range) in the pulsar magnetosphere by current sheet reconnection
\citep{2015MNRAS.448..606C}, broadly consistent with $E_{\rm min}$ we obtained.
A fraction of these particles will then undergo further acceleration at the
termination shock, while those that are not efficiently accelerated remain in
the low-energy population and manifest as the observed steepening at the GeV
scale. Note that simulations predict an additional population of very low-energy
particles with $\gamma \sim 10$. These are not detectable with our observations
since they emit at very low frequency with small radiation power compared with
those $\gamma\sim10^3$ particles.\footnote{Both the synchrotron characteristic
frequency and radiation power scale as $\gamma^2$.} 

Alternatively, the low-energy steepening could arise naturally from relativistic
dissipation processes in PWNe. PIC simulations of various particle acceleration mechanisms
%, including DSA and shock-driven magnetic reconnection (\textbf{different from current sheet reconnection, emphasize it or just remove "including ...."}),
commonly find a thermal
(Maxwellian-like) component at low energy and a PL or broken PL tail at high
energy \citep{Sironi&Spitkovsky2009,ss11a,ss11b}. The Maxwellian component results from bulk energy dissipation and subsequent thermalization of particles, while the PL is only contributed by particles that are able to escape
upstream and undergo multiple shock crossings \citep{2009ApJ...698.1523S}. The
post-shock particle distribution depends on the ratio of stripe wavelength to
Larmor radius and magnetization \citep{2007A&A...473..683P,ss11a}.

The injection break inferred from our modeling, $E_{\rm min}\approx1-4$\,GeV, is
four orders of magnitude above the characteristic Maxwellian peak energies reported in PIC
simulations of the order of $\sim$0.1\,MeV. However, the comparison is not
straightforward. The injection energy scale in PIC simulations depends
sensitively on the upstream wind parameters, particularly the magnetization and
the bulk Lorentz factor ($\gamma_0\sim10$ in the simulations). A GeV-scale
break, as inferred for the Mouse, could merely reflect specific wind conditions,
such as relatively low magnetization at the termination shock.

The ability to measure the lower-energy part of the injection spectrum (either a
cutoff or a Maxwellian component) could provide stringent constraints on
the particle multiplicity. Specifically, integrating over the entire energy
distribution of the particles yields an average Lorentz factor $\gamma_{\rm
mean} \simeq 2.6\times 10^4$, corresponding to a multiplicity defined as $\mu =
E_{\rm max} / (\gamma_{\rm mean} m_e c^2) \simeq 2.5 \times 10^5$, for $B_{\rm
pwn}\sim50\,\mu$G. This is consistent with predictions from pair creation models
\citep{2015ApJ...810..144T}, and does not depend on the distribution models,
indicating that either a sharp low-energy cutoff or a smoother Maxwellian
low-energy tail does not modify the overall picture.

\begin{table*}[ht!]
\centering
\caption{PWNe with direct observations or indications of a spectral break that is potentially caused by the low-energy steepening in injected spectra: \label{tab:sources}}
\begin{tabular}{lcccc}
\toprule
\multirow{2}{*}{PWN} & \multirow{2}{*}{Spectral break (GHz)} & \multirow{2}{*}{Inferred $E_{\rm min}$ (GeV)} & \multirow{2}{*}{Evolutionary stage} & \multirow{2}{*}{Ref.}\\
 &  & & & \\
\hline
\multicolumn{4}{l}{\textbf{Spectral break directly observed}} \\
\hline
Mouse & 3.7 & 1.4--4.0 & Bow shock & This work\\
Boomerang & 4.3 & 0.4--22.3 & Reverse shock interaction  & (1), (2)\\
CTB 80 & 4--5 & \nodata & Young PWN in SNR? & (3), (4)\\
Vela & $\sim5.2$ & \nodata & Reverse shock interaction & (5)\\
\hline
\multicolumn{4}{l}{\textbf{Spectral break estimated from multi-wavelength SED}} \\
\hline
Kes 75 & 0.1--1 & 2 & Young PWN in SNR  & (6)\\
Vela X& 10--100 & $\sim3$ & Reverse shock interaction & (7) \\
G21.5$-$0.9 & 10--100 & 12.5 & Young PWN in SNR & (8) \\
G11.2$-$0.3 & $\sim$100 & \nodata & Young PWN in SNR & (9) \\
\hline
\end{tabular}
\vspace{2mm}
\footnotesize
\begin{flushleft}
References: (1) \citealt{2006ApJ...638..225K}; 
(2) \citealt{2024ApJ...960...75P}; 
(3) \citealt{2003AJ....126.2114C}; 
(4) \citealt{2005AA...440..171C}; 
(5) \citealt{2003MNRAS.343..116D};
(6) \citealt{2021ApJ...908..212G}; 
(7) \citealt{2011ApJ...743L..18M}; 
(8) \citealt{2020ApJ...904...32H}; 
(9) \citealt{2025ApJ...988..163Z}
\end{flushleft}
\end{table*}

\begin{comment}
The relativistic Larmor radius is 
\begin{equation}
R_L\equiv\frac{E}{eB_{\rm pwn}}~.
\end{equation}
The Larmor radius corresponding to the low-energy steepening in best-fit Maxwellian-PL model ($R_{L,E_{\rm min}}$) is $3.3\times10^{11}$\,cm.

In relativistic pulsar winds, the plasma skin depth ($\delta_e$) characterizes the microscopic length scale over which electromagnetic (EM) fields can penetrate the plasma before the plasma collectively responds to shield them. At scales exceeding the plasma skin depth, the detailed microphysics (particle scattering, isotropization) are “hidden”. Collective plasma processes, including shocks and magnetic reconnection, occur at scales larger than the plasma skin depth.
%The termination shock’s width is typically a few times the upstream skin depth. 
The skin depth is calculated as $\delta_e=c/\omega_p$, where $\omega_p$ is the plasma frequency,
\begin{equation}
\omega_p=\sqrt{\frac{4\pi n_ee^2}{\gamma m_e}}~,
\end{equation}
with $\gamma$ being the mean Lorentz factor of the injected plasma wind and $n_e$ being the number density of upstream plasma measured in the lab frame. For PWN, $n_e$ at the termination shock is
\begin{equation}
n_e=\frac{\dot{E}}{4\pi R_{\rm ts}^2\gamma m_ec^3}~,
\end{equation} where the termination shock radius $R_{\rm ts}\approx0.018$\,pc is given by \cite{gvc+04}.

Adopting $\dot{E}=2.5\times10^{36}$\,erg/s, $\gamma=10^4$, the plasma skin depth $\delta_e$ is $1.1\times10^{11}$\,cm, comparable to the relativistic Larmor radius of the low energy $R_{L,E_{\rm min}}$. 
\end{comment}

\subsection{Low-Frequency Spectral Breaks in Other PWNe}
In addition to the Mouse, a number of young PWNe also have spectral breaks
observed or inferred in the radio band. Their properties are listed in
Table~\ref{tab:sources}. The Boomerang shows a break at $\sim$4.3\,GHz
\citep{2006ApJ...638..225K}, comparable to that of the Mouse. This is attributed to either synchrotron cooling with a strong $B$-field of 2.6\,mG or a minimum
injected particle energy ($E_{\rm min}$) between 0.4 and 22.3\,GeV
\citep{2024ApJ...960...75P}. Similarly, the PWN inside CTB 80 shows a hint
of a spectral turnover at a few GHz, but this is interpreted as absorption by ionized gas in the surrounding SNR \citep{2005AA...440..171C}. Likewise, the Vela nebula, currently in the reverse-shock interaction phase, also displays a similar turnover at $\sim$5.2\,GHz in its flux density measurements \citep{2003MNRAS.343..116D}.
For other sources in Table~\ref{tab:sources}, the evidence is less clear due to
poor frequency coverage in radio. Their breaks are inferred from broadband
modeling of the SED from radio to X-ray bands
\citep[e.g.,][]{baa11,2011ApJ...743L..18M}.

We stress that in Table~\ref{tab:sources} the Mouse is the only bow-shock PWN
and all others are young sources inside SNRs, such that their spectra could be
complicated by environmental effects, e.g., absorption or different particle
populations. Also, the Mouse is the only case that shows a rising spectrum below
the break, giving strong evidence for the steepening of the particle
distribution at low energy. To investigate if this is a common phenomenon among
bow-shock PWNe, more observations with better frequency coverage will be needed.

\subsection{Overall Injected Particle Spectrum}
% As mentioned in Section~\ref{sec:intro}, conventional first-order Fermi acceleration and shock-driven magnetic reconnection predict a single PL and a broken PL, respectively, while relativistic PIC simulations incorporating Weibel instability found a Maxwellian-PL distribution. 
We showed that the SED of the Mouse from radio to X-ray bands can be well
modeled by synchrotron emission from a single population of particles with
cooling taken into account. Remarkably, a single PL particle distribution is
sufficient to reproduce the overall broadband spectrum above the break. This is
no surprise since an extrapolation of the observed radio spectrum
$\alpha_2=0.51$ to the X-ray band gives a photon index
$\Gamma_X=\alpha_2+1.5\approx2$ after cooling, which well agrees with the
\emph{Chandra} result of $\Gamma_X=2.09$. The best-fit particle PL index of $p =
2.17$ is generally compatible with theoretical expectations and simulation
results based on the conventional DSA scenario \citep[e.g.,
$p\approx2.2$--$2.3$;][]{2001MNRAS.328..393A}. For the Maxwellian-PL model, the
main distinction is that it offers a more physical description of the low-energy
cutoff, as discussed in Section~\ref{sec:physics} above, while the high-energy
part is the same.

As Figure~\ref{fig:SED} indicates, the existing data cannot distinguish between the three distribution functions we tried (PL, broken PL, and Maxwellian-PL).
In the future, precise measurements spanning the submillimeter to infrared range, enabled by facilities like the Very Large Telescope or the James Webb Space Telescope, will be highly beneficial. Even if a simple PL or a Maxwellian-PL provides a statistically slightly preferred fit, a broken PL is also suggested by theories. The shock-driven magnetic reconnection process can generate low-energy particles with a hard spectrum of $p_0 \sim 1.5$
\citep{ss11b}. The latter can vary from 1.5 to 2, depending on the pulsar wind magnetization \citep{ss14}. On the other hand, the high-energy component is commonly attributed to DSA, same as the PL case above. All these are consistent with the best-fit broken PL parameters we obtained ($p_0 = 1.47$ and $p = 2.17$).

Note that the broken-PL model is also commonly adopted in SED modeling of young PWNe \citep[e.g.,][]{baa11,torres14}. These studies found similar parameters to our results, for instance, $p_0\sim$1.6 for the Crab Nebula \citep{2015MNRAS.451.3145Z}, $\sim$1.2 for 3C~58, and $\sim$1.7 for Kes 75 \citep{baa11}. Similarly, the reported ranges of $p$ and $E_{\rm bre}$ are also
comparable to ours (e.g., from $\sim 20$\,GeV for MSH 15$-$5\emph{2} to $\sim 400$\,GeV for the Crab). These provide a clear indication that, despite the significantly smaller size of the system (the distance from the pulsar to the termination shock in the Mouse PWN is about an order of magnitude smaller than
in the Crab, 30 times smaller when accounting for the difference in the light cylinder size), particle acceleration might occur in a fundamentally similar way. These findings also disfavour alternative origins for radio-emitting particles, such as fossil populations, entrainment, or turbulent
bulk acceleration, leaving shock-driven magnetic reconnection as the most natural explanation.

Finally, all our models suggest $B_{\rm pwn}\sim50\,\mu$G,
weaker than the equipartition field of 250\,$\mu$G previously
reported \citep{kkp+18}. Resolving this ambiguity will likely require constraints from optical wavelength observations.

%}%%%%%%%%%%%%%%%%%%%%%%%%%%%%%%%%%%%%%%%%%%%%%%%%%%%%%%%%%%%%%%%%%%%%%%%%%%%%%%%%%%%%
\section{Summary} 
\label{sec:Summary}
In this paper, we report the broadband spectrum of the Mouse PWN obtained from
multiwavelength observations. The result clearly indicates a spectral break at
$3.7^{+0.7}_{-0.6}$\,GHz. We argued that this is likely an intrinsic feature of
the injected particle distribution, rather than synchrotron cooling, synchrotron
self-absorption, or free-free absorption effects. We showed that the broadband SED
is well modeled by synchrotron emission from a single particle population,
with either a PL with a low-energy cutoff, a broken PL with a low-energy cutoff,
or a Maxwellian plus PL energy distribution. The best-fit model parameters are
generally compatible with theory and simulation results. We discussed possible
physical origins of such a cutoff, including the characteristic energy of
leptons leaving the pulsar magnetosphere and energy dissipation after the
particles cross the termination shock.

This work demonstrates that detailed observations and modeling of SED provide a
powerful probe of the underlying particle distribution. For further
investigation, more observations of similar bow-shock systems will be needed to
determine if the spectral break found in the Mouse is ubiquitous among PWNe. The
results will provide valuable constraints for testing and refining particle
acceleration theories.

\begin{acknowledgments}
We thank Duncan Campbell-Wilson for providing the MOST image and Rahul Basu for the GMRT data.
We are grateful to Yihan Liu for performing the ATCA observations in 2023. We
thank Bing Zhang and Farhad Yusef-Zadeh for the useful discussion.  Z.S.\ and
C.-Y.N.\ are supported by a GRF grant of the Hong Kong Government under HKU
17303221. Z.S.\ is supported by the Dissertation Year Fellowship of the
University of Hong Kong.

This paper makes use of the following ALMA data:
ADS/JAO.ALMA\#2025.1.01325.S and ADS/JAO.ALMA\#2019.1.01187.S\@. ALMA is a partnership of ESO (representing its
member states), NSF (USA) and NINS (Japan), together with NRC (Canada), NSTC
and ASIAA (Taiwan), and KASI (Republic of Korea), in cooperation with the
Republic of Chile. The Joint ALMA Observatory is operated by ESO, AUI/NRAO and
NAOJ.
%% TODO(ALMA): verify the partnership sentence verbatim against
%%   https://almascience.org/alma-data/publication-acknowledgement
%%   -- ALMA revises the partner list, so copy the current text rather than this one.
\end{acknowledgments}

\appendix
\section{Details of Individual Observation}

\subsection{ATCA}
\label{Appendix:ATCA}
The ATCA observations were carried out on 2023 April 1 (L/S band), May 13 (C/X band), and September 10 (K band), using the 750C, EW352, and H168 array configurations, respectively.
The data were obtained with the CABB backend, centered at 2.1, 5.5, 9.0, 16.7, and 21.2\,GHz, each with a bandwidth of 2\,GHz. PKS~1934$-$638 was used as the primary flux calibrator, while PKS~1741$-$312 and PKS~1253$-$055 were adopted as the phase and bandpass calibrators.

\begin{figure}[hbt]
\centering
\includegraphics[width=0.49\textwidth]{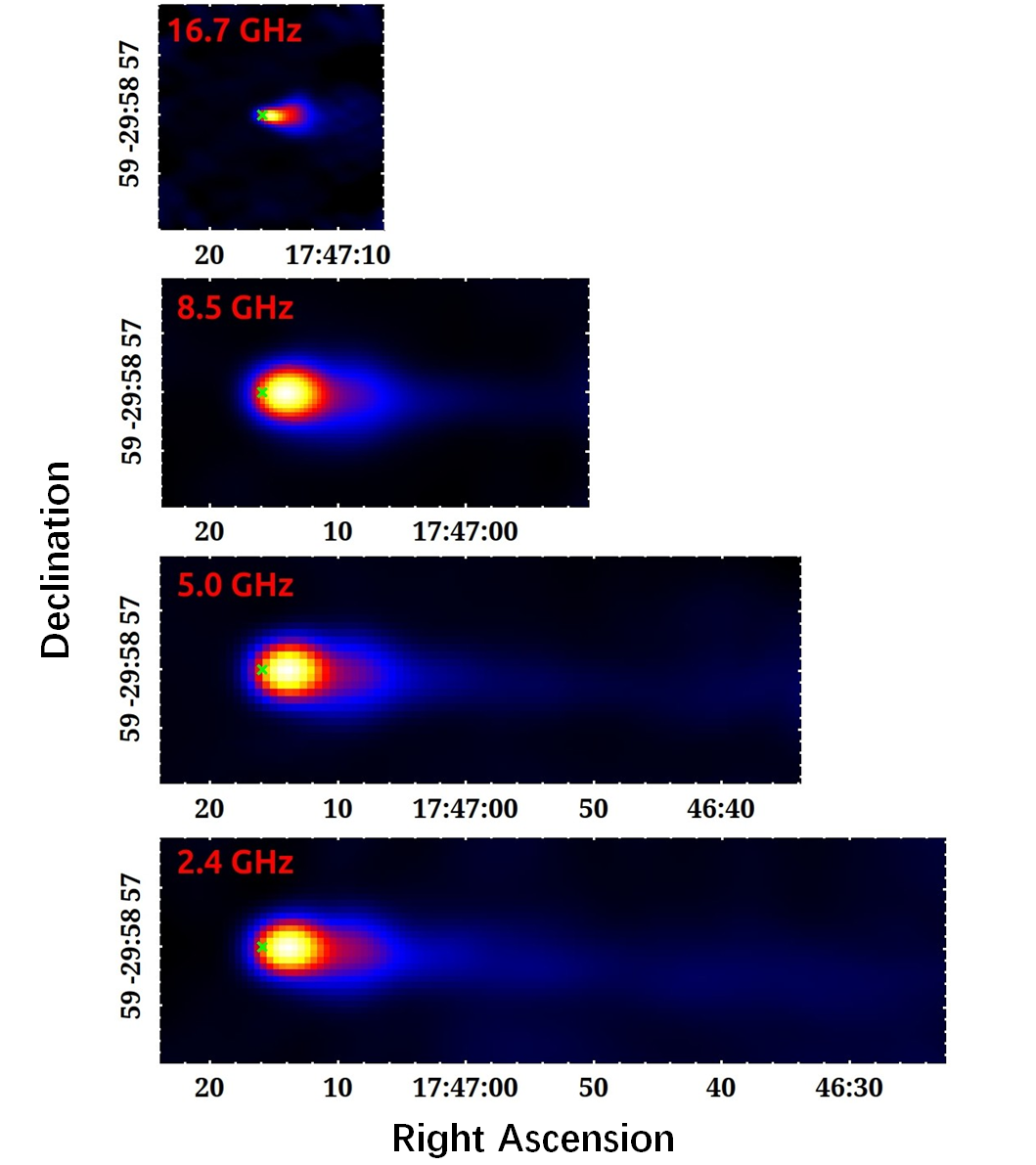}
\caption{ATCA total intensity maps of the Mouse at 16.7, 8.5, 5.0, and 2.4\,GHz. The green cross marks the pulsar position.
\label{fig:ATCA}}
\end{figure}

Data reduction was performed using the {\tt MIRIAD} package following standard ATCA procedures.
Antenna~6, which has the longest baseline relative to the other antennas, was excluded to avoid large gaps in the $uv$ coverage.
The L/S and C/X band data were divided into four sub-bands, while the K-band data were divided into two sub-bands.
Radio frequency interference (RFI) was identified and flagged using a combination of automatic and manual flagging with \texttt{pgflag}.
The sub-band centred at 1.3\,GHz is severely affected by RFI and was therefore excluded from the subsequent analysis. Flux density calibration was performed using the standard ATCA flux scale, and complex gain calibration was derived from regular observations of the phase calibrator.
Imaging was carried out using \texttt{invert} with Briggs visibility weighting (robust = 1), and the images were deconvolved using the \texttt{maxen} algorithm.
Since our primary goal is to measure the total flux density rather than resolve fine-scale structures, a relatively large Gaussian $uv$ taper was applied during imaging to suppress sidelobe noise and improve the robustness of the integrated flux density measurements. The resulting ATCA maps of the Mouse at 16.7, 8.5, 5.0, and 2.4\,GHz are shown in Figure~\ref{fig:ATCA}.

\subsection{ALMA}
\label{Appendix:ALMA}
The Mouse was observed with the Atacama Compact Array (ACA) 7\,m array under
project 2025.1.01325.S (PI: Shi). Three executions were carried out: Band~1 on
2025 October 24, Band~4 on 2026 April 20, and Band~3 on 2026 June 22. The source was covered by mosaics of 5, 23, and 26 pointings in Bands~1, 4,
and 3, respectively.

Each execution used four spectral windows in continuum mode. Each window has
an effective bandwidth of $\sim1.7$\,GHz, giving an aggregate bandwidth of
$\sim6.8$\,GHz. The windows are centred at 37.2, 39.1, 41.2, and 43.1\,GHz in
Band~1, at 90.5, 92.5, 102.5, and 104.5\,GHz in Band~3, and at 138.0, 140.0,
150.0, and 152.0\,GHz in Band~4.

We used the calibrated and imaged products delivered by the ALMA pipeline
(versions 2025.1.0.35 and 2025.1.0.37, running under {\tt CASA}~6.6.6). The
images were produced with \texttt{tclean} using the mosaic gridder and the
multi-term multi-frequency synthesis deconvolver with $\mathtt{nterms}=2$. The
visibilities were weighted with the Briggs scheme at $\mathtt{robust}=0.5$, and
the clean masks were generated with \texttt{auto-multithresh}. Phase
self-calibration was applied, and all images are corrected for the primary-beam
response. The aggregate-continuum synthesized beams are
$40\farcs2\times22\farcs9$, $19\farcs6\times8\farcs4$, and
$10\farcs4\times7\farcs0$ in Bands~1, 3, and 4. The rms noise of the individual
spectral-window images ranges from 2.0 to 2.8\,mJy\,beam$^{-1}$ in Band~1, from
1.2 to 1.6\,mJy\,beam$^{-1}$ in Band~3, and from 0.7 to 0.8\,mJy\,beam$^{-1}$ in
Band~4.

\begin{figure*}[hbt]
\centering
\includegraphics[width=0.99\textwidth]{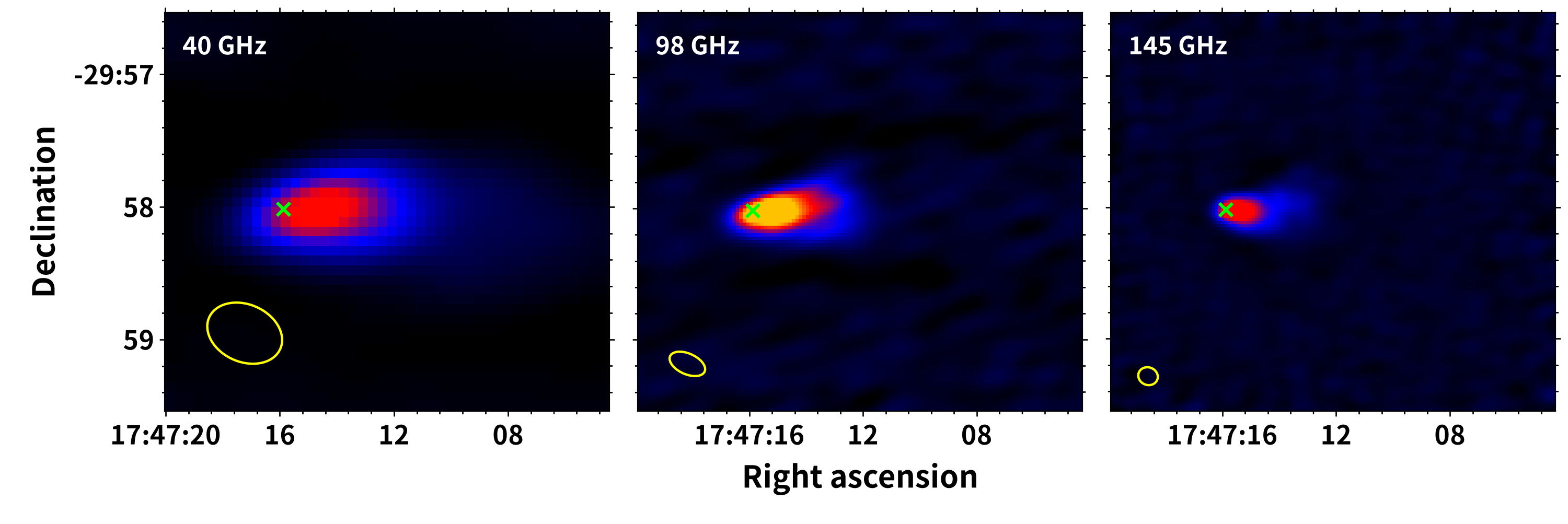}
\caption{ALMA images of the Mouse at 40, 98, and 145\,GHz. The green cross marks the position of the pulsar,
and the yellow ellipse in the lower left of each panel shows the synthesized
beam.
\label{fig:ALMA}}
\end{figure*}

The 7\,m array alone does not sample the shortest spacings, and no
total-power data were taken. Emission that is smooth on scales larger than the
maximum recoverable scale of the array is therefore filtered out. We accordingly
treat the integrated flux densities from these data as lower limits
(Section~\ref{sec:observation}). The aggregate-continuum images of the three
bands are shown in Figure~\ref{fig:ALMA}.

We further retrieved an archival ALMA 12\,m array observation of the Mouse at
100\,GHz, taken on 2019 December 12 in the C43-1 configuration (project
2019.1.01187.S; PI: Klingler). As for the ACA data, the integrated flux density
from this dataset is treated as a lower limit.

\subsection{IRAM 30m Telescope}
\label{Appendix:IRAM}
\begin{figure*}[hbt]
\centering
\includegraphics[width=0.99\textwidth]{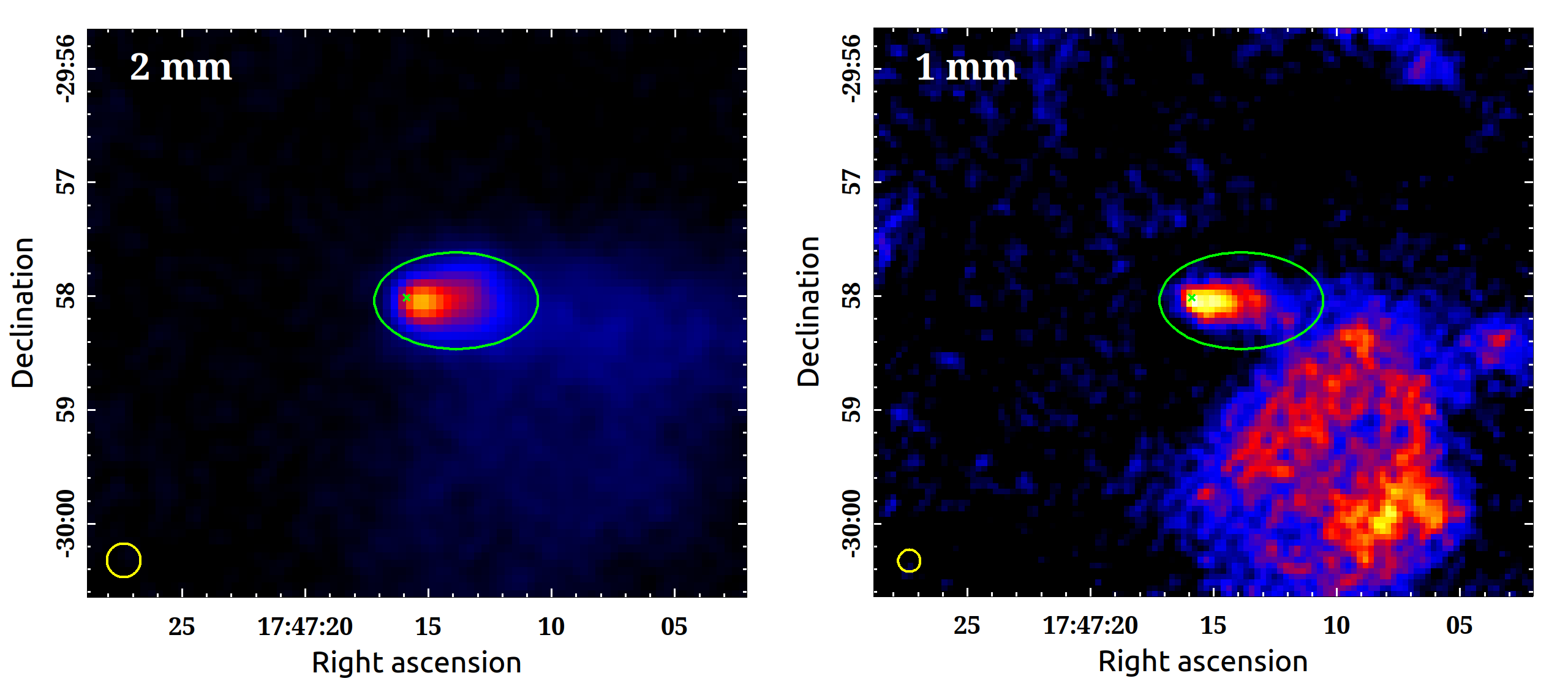}
\caption{IRAM 30m telescope observations of the Mouse with NIKA2 camera at 2\,mm (left) and 1.15\,mm (right) wavelengths. The yellow circles denote the beam size. The green ellipse indicates the PWN region, and the cross denotes the position of the pulsar. \label{fig:IRAM}}
\end{figure*}

In the millimeter band, continuum observations of the Mouse PWN were taken with the dual-band camera NIKA2 of the IRAM 30m telescope on 2025 October 24. A total of 35 valid scans were acquired on the target region and the data were reduced using the official pipeline \texttt{PIIC} following the standard science workflow. The default reduction templates applied were \texttt{mapTP\_a2\_template.piic} and \texttt{mapTPoptionalSets.piic}.
Two reductions were performed: the first reduction, with zero pipeline iterations, was conducted primarily to verify data integrity. Following the establishment of the data associated files (DAFs) for calibration, a second reduction was carried out employing 20 pipeline iterations. This process yielded final coadded maps for both NIKA2 bands: 2\,mm map from the Array 2 and 1.15\,mm map from the combined Array 1 and 3. The achieved RMS noise levels are approximately 0.9\,mJy\,beam$^{-1}$ at 2\,mm and 1.9\,mJy\,beam$^{-1}$ at 1.15\,mm, respectively. The flux density measurements and spectral analyses presented in this work are based on the final 20-iteration coadded map, see Figure~\ref{fig:IRAM}.
\subsection{JCMT}
\label{Appendix:JCMT}
JCMT SCUBA-2 850 and 450 $\mu$m observations were taken on 2025 July 12 to 14 with an allocated time of 8 hours in 15-arcmin Pong mode.
The data were reduced using the automated \texttt{ORAC-DR} pipeline within the {\tt STARLINK} software environment, following standard SCUBA-2 data reduction procedures.

\begin{figure*}[hbt]
\centering
\includegraphics[width=0.99\textwidth]{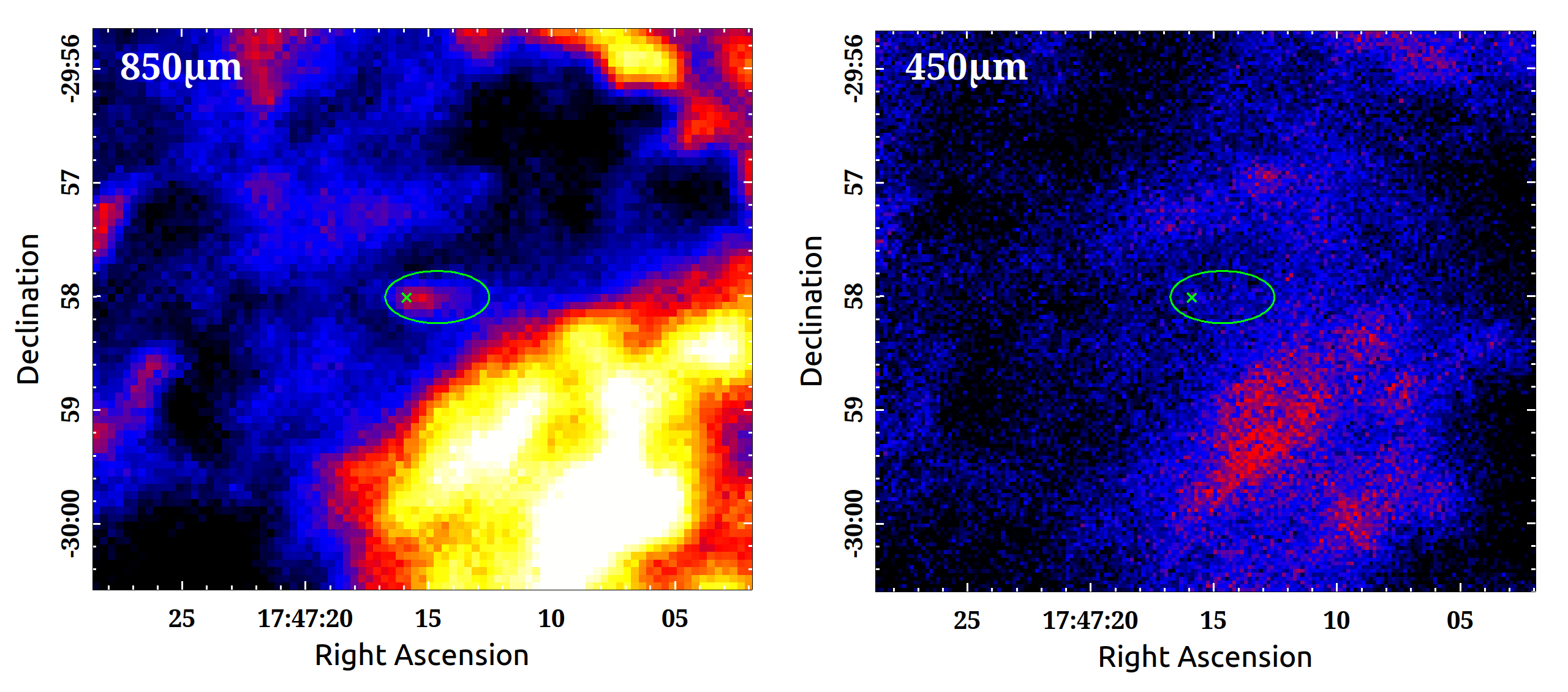}
\caption{JCMT 850\,$\mu$m (left) and 450\,$\mu$m (right) wavelength observations of the Mouse. The ellipse indicates the PWN region, and the cross denotes the position of the pulsar. Due to the non-detection in the 450\,$\mu$m band, the flux density measurement is treated as an upper limit. \label{fig:JCMT}}
\end{figure*}

The raw data were processed using the default SCUBA-2 pipeline setup, with map-making performed using the {\tt SMURF} task \texttt{makemap}, adopting the standard JCMT configuration file \texttt{dimmconfig\_jsa\_generic.lis}.
The data were reduced using the \texttt{REDUCE\_SCAN\_CHECKRMS} recipe, which performs iterative map-making, noise estimation, and quality control.
The individual reduced maps were then combined to produce the final mosaic image. To enhance the detectability of compact emission, a matched filter was applied to the final mosaic using standard SCUBA-2 procedures.
The resulting JCMT observations are shown in Figure~\ref{fig:JCMT}.

\subsection{Spitzer} \label{Appendix:Spitzer}
\begin{figure*}[htp]
	\centering
	\includegraphics[width=0.75\textwidth]{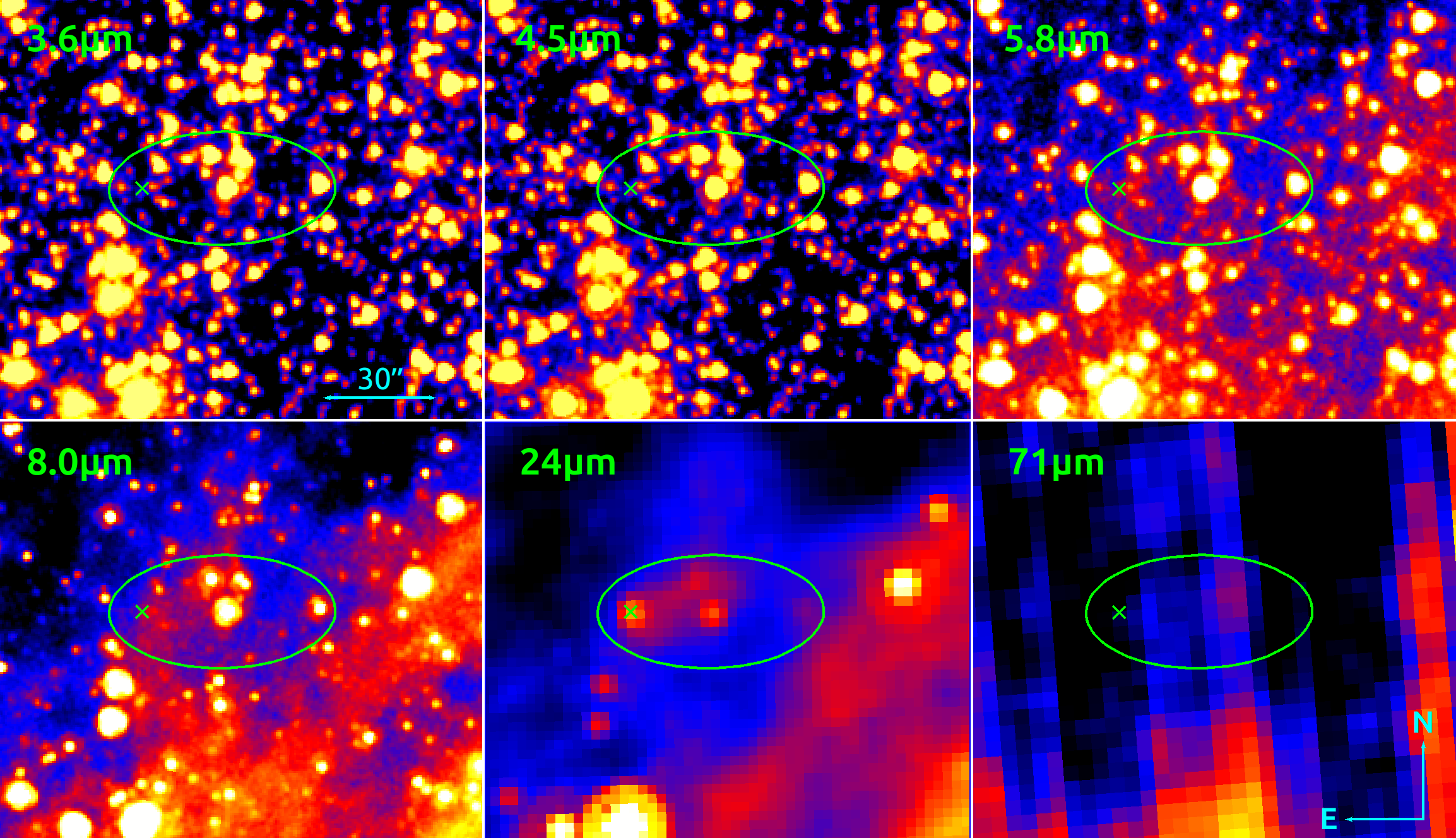}
	\caption{\emph{Spitzer} infrared images of the Mouse taken at 3.6, 4.5, 5.8, 8.0\,$\mu$m with IRAC and MIPSGAL 24, 70\,$\mu$m.
	The cross indicates the position of \psr\ and the ellipse of $1\arcmin\times30\arcsec$ shows the extraction region used for estimating the flux upper limits.}
	\label{fig:Spitzer}
\end{figure*}

The infrared data employed in this work are obtained from the archive of the \emph{Spitzer} Space Telescope. We retrieved Level 1 Basic Calibrated Data (BCD) and Level 2 Post-BCD (PBCD) products from the \emph{Spitzer} Science Archive, which were preprocessed via the standard pipeline of the \emph{Spitzer} Science Center (SSC). BCD products include corrections for instrumental artifacts, comprising dark current subtraction, flat-fielding, linearity correction, and absolute flux calibration; PBCD products are fully calibrated, co-added mosaics produced by the SSC pipeline, which we adopt directly for our scientific analysis. Final \emph{Spitzer} images at 3.6, 4.5, 5.8, and 8.0\,$\mu$m (taken with IRAC) and 24 and 70\,$\mu$m (from MIPS) are presented in Figure~\ref{fig:Spitzer}.

\subsection{NuSTAR and Chandra \label{Appendix:Xray}}
\begin{figure}[hbt]
\centering
\includegraphics[width=0.49\textwidth]{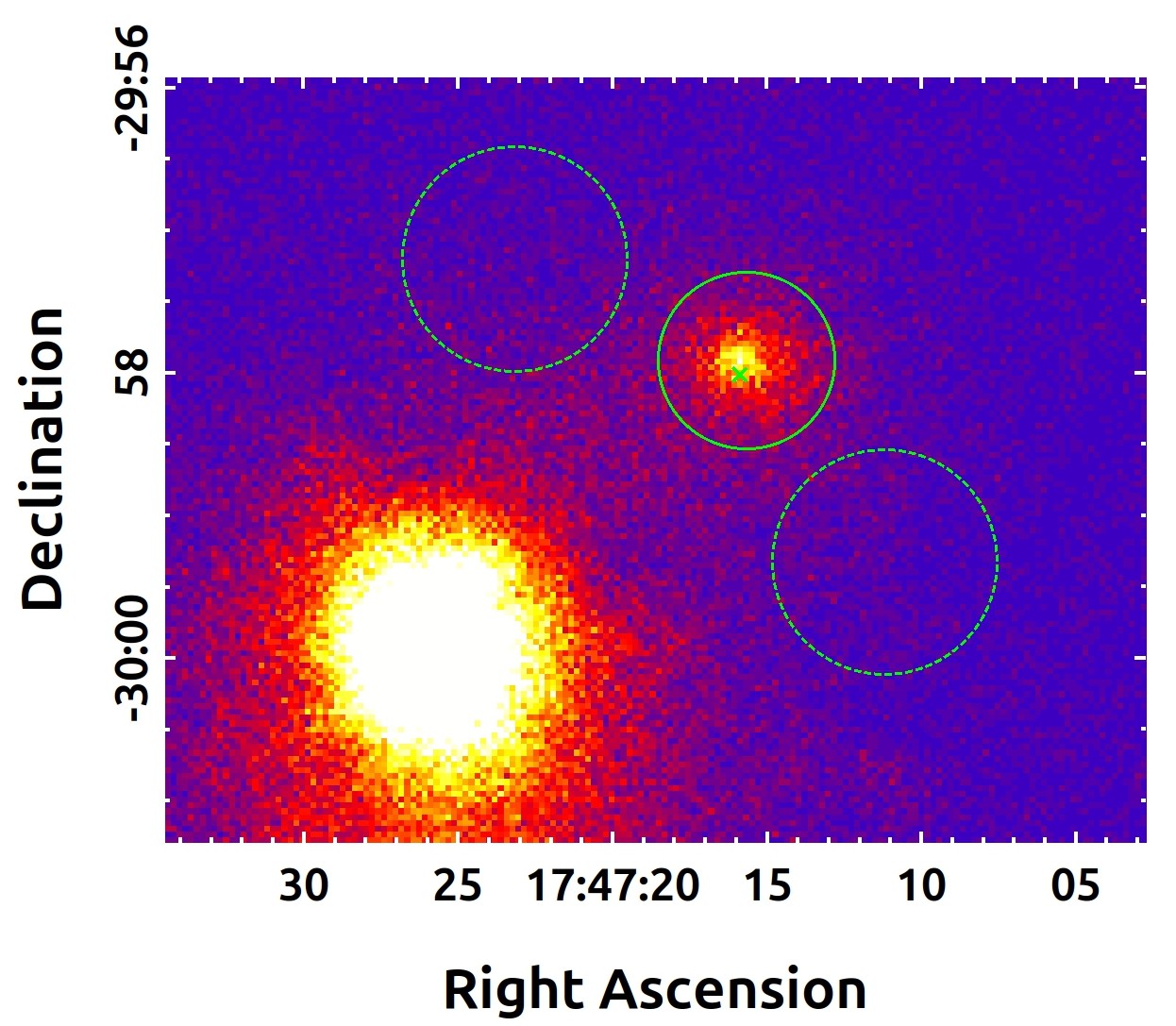}
\caption{\nus\ FPMA image of the Mouse in the 3--30\,keV band. The solid line indicates the 50\arcsec\ radius aperture for the extraction of the PWN spectrum.
The background extraction regions are indicated by the dashed lines.
All these regions have the same distance from SLX 1744$-$299, the bright source to the southeast.\label{fig:nustar}}
\end{figure}

In the X-ray band, we performed a joint spectral fit using the \cxo\ and \nus\ data (see below for more details). Owing to the limited angular resolution of \nus, the pulsar emission cannot be spatially resolved and reliably subtracted (see Figure~\ref{fig:nustar}); therefore, the joint-fit results include a contribution from the pulsar and are treated as upper limits on the nebular emission. For our SED modeling, we adopt the spatially resolved results for the entire X-ray tail from \citet{kkp+18}, in which the pulsar emission was explicitly excluded.
The X-ray spectrum of the entire PWN tail is characterised by a photon index of $\Gamma_X = 2.09 \pm 0.02$ and a PL normalization at 1\,keV of $\mathcal{N}_{-4} = (16.2 \pm 0.03)\times10^{-4}$\,photons\,s$^{-1}$\,cm$^{-2}$\,keV$^{-1}$, see Figure~\ref{fig:SED}. 

Here we show the joint-spectral fit results of the \cxo\ and \nus\ data. The Mouse was observed with \nus\ on 2018 Jul 29 with an exposure time of 40.4\,ks (ObsID 30401036002).
We carried out data reduction using NuSTARDAS v2.0.0 in \texttt{Heasoft v6.28} with calibration database (CALDB) version 20210104.
The data were first processed using the standard pipeline \texttt{nupipeline}.
The pulsar and PWN system is clearly detected up to $\sim$30\,keV,
but the spatial resolution of \nus\ is not sufficient to resolve the nebular structure.
Figure~\ref{fig:nustar} shows an image in the 3--30\,keV band from FPMA.
The FPMB image is very similar.
To search for pulsations, we applied barycentric corrections to the photon arrival times, then extracted counts from a 30\arcsec\ radius aperture centered on the pulsar.
We folded the light curve according to the pulsar ephemeris obtained from a timing program for \emph{Fermi},
but found no pulsations in the 3--30\,keV band.

For spectral analysis, we extracted source counts from a 50\arcsec\ radius aperture enclosing the entire PWN as seen in the \cxo\ X-ray image.
This aperture is expected to enclose $\sim$70\% energy of a point source.
We avoid using a too large extraction region as there is a bright X-ray source, SLX 1744$-$299,
3\arcmin\ southeast of the PWN (see Fig.~\ref{fig:nustar}).
Based on a model PSF, we estimated that this source only contributes about 10 counts in 3--30\,keV in our extraction aperture; it is therefore negligible.
Nonetheless, we chose background regions that have the same separation (3\arcmin) from SLX 1744$-$299 as the PWN.
The source and background regions are shown in Figure~\ref{fig:nustar}.
We also tried an annular background region centered on the PWN and the results are very similar.
After background subtraction, we obtained 5150 and 5050 source counts in 3--30\,keV from FPMA and FPMB, respectively.
We also included \cxo\ spectrum in the analysis as they provide a better constraint on the absorption column density (\nh).
We used the same data (ObsIDs 2834, 14519, 14520, 14521, and 14522) as presented by \citet{kkp+18} and extracted the
overall PWN spectrum from a region enclosing the entire nebula \citep[see Figure~3 in][]{kkp+18}.

The spectral fitting was performed with the \texttt{Sherpa} package.
We grouped the \nus\ and \cxo\ spectra to a minimum of 60 and 75 counts per bin, respectively,
and restricted the energy ranges to 3--30\,keV and 0.5--8\,keV, respectively.
We first fit the \cxo\ spectrum using a simple PL with the \citet{wam00} ISM absorption model.
This gives the best-fit $N_{\rm H}=3.96\times10^{22}\,\mathrm{cm}^{-2}$ and photon index $\Gamma=1.82$.
We then fixed \nh\ at this value and fit the PL to the \nus\ spectrum.
This yields a slightly larger $\Gamma=1.98$, indicating a softer spectrum.
Next, we carried out a joint analysis to both spectra.
We fixed \nh\ and allowed a constant scaling factor between the \nus\ and \cxo\ flux to account for different encircled energy fractions of the extraction regions and any cross-calibration uncertainty.
A simple PL fit gives $\Gamma=1.90$, exactly the mean of the two individual fits above.
We also tried a log-parabolic (logpar) model,
since there is evidence of spectral softening in hard X-rays.
The mathematical form is given by
\begin{equation}
f(E)=\left(E/E_0\right)^{-\alpha-\beta\log(E/E_0)}~,
\end{equation}
where the reference energy $E_0$ is fixed at 1\,keV,
$\alpha$ is the photon index at $E_0$, and $\beta$ is the curvature term.
We obtained $\alpha=1.70$ and $\beta=0.13$.
In terms of reduced $\chi^2$ value, the logpar model is better than the PL and an $F$-test suggests that the improvement is significant.
The best-fit parameters of all fits with 1$\sigma$ uncertainties are listed in Table~\ref{tab:xray_fits}.

\begin{deluxetable*}{ccccccccc}[thb]
\tablecaption{Best-fit \cxo\ and \nus\ X-ray Spectra of the Mouse \label{tab:xray_fits}}
\tablehead{\colhead{Telescope} & \colhead{Model} & \colhead{\nh} & 
\colhead{$\Gamma$} & \colhead{$\alpha$} & \colhead{$\beta$} &
\colhead{$F^{\rm abs}_{\rm 0.5-8\,keV}$} &
\colhead{$F^{\rm abs}_{\rm 3-30\,keV}$}& \colhead{$\chi^2_\nu/\nu$}\\
&& ($10^{22}$\cms) &&&& \colhead{(\ergcm)} & \colhead{(\ergcm)}}
\startdata
\cxo & PL & $3.96\pm0.06$ & $1.82\pm0.03$ & \nodata & \nodata & $5.6\times10^{-12}$ & \nodata & 1.11/466\\ 
\nus & PL & $3.96^\ast$ & $1.98\pm0.03$ & \nodata & \nodata & \nodata & $7.0\times10^{-12}$ & 1.34/179\\ 
\cxo+\nus & PL & 3.96\tablenotemark{$\ast$} & $1.90\pm0.01$ & \nodata & \nodata & $5.5\times10^{-12}$ & $7.2\times10^{-12}$  & 1.22/647\\
\cxo+\nus & logpar & $3.96^\ast$ & \nodata & $1.70\pm0.04$ & $0.13\pm0.03$ &
$5.5\times10^{-12}$ & $7.0\times10^{-12}$ & 1.19/646
\enddata
\tablecomments{all uncertainties are at 1$\sigma$ confidence level. The 0.5--8\,keV and 3--30\,keV absorbed fluxes are obtained from the fitting of the \cxo\ and \nus\ data, respectively.}
\tablenotemark{$\ast$}{held fixed during the fit.}
\end{deluxetable*}

\bibliography{references,ncy}
\bibliographystyle{aasjournalv7}

\end{document}